\documentclass[12pt,a4]{article}
\usepackage{amsmath}
\usepackage{amsfonts}
\usepackage{hyperref}

\newcommand {\be}{\begin{equation}}
\newcommand {\ee}{\end{equation}}
\newcommand {\nn}{\nonumber}
\newcommand {\pd}{\partial}
\newcommand {\ad}{{d^\prime}}
\newcommand {\Verma}[3]{D\left(#1,#2\right)_{o(#3,2)}}
\newcommand {\Vol}[1]{d\mu_{#1}\:}
\newcommand {\ie}{{\it i.e.,\ }}

\newenvironment{prf}{%
\noindent{\sc Proof.}}%
{\noindent{\hfill\mbox{\rule{.5em}{.5em}}\,}\par\medskip}
\newtheorem{Thr}{Theorem}
\newtheorem{Cor}{Corollary}

\begin{document}
\begin{flushright}
\vspace{1mm}
FIAN/TD/21-08\\
\end{flushright}

\vspace{1cm}

\begin{center}
{\bf \Large  On Dimensional Degression in $AdS_d$} \vspace{1cm}

\textsc{A.~Yu.~Artsukevich\footnote{artsukevich@lpi.ru} and
M.~A.~Vasiliev\footnote{vasiliev@lpi.ru}}

\vspace{.7cm}

{\em I.E.Tamm Department of Theoretical Physics, P.N.Lebedev
Physical Institute,\\
Leninsky prospect 53, 119991, Moscow, Russia}

\vspace{3mm}

\end{center}

\vspace{2cm}

\abstract{We analyze the pattern of fields in $d+1$ dimensional
anti-de Sitter space in terms of those in $d$ dimensional anti-de
Sitter space. The procedure, which is neither dimensional reduction
nor dimensional compactification, is called dimensional degression.
The analysis is performed group-theoretically for all totally
symmetric bosonic and fermionic representations of the anti-de
Sitter algebra. The field-theoretical analysis is done for a massive
scalar field in $AdS_{d+\ad}$ and massless spin one-half, spin one,
and spin two fields in $AdS_{d+1}$. The mass spectra of the
resulting towers of fields in $AdS_d$ are found. For the scalar
field case, the obtained results extend  to the shadow sector those
obtained by Metsaev in \cite{Metsaev:2000qb} by a different method.}

\newpage
\tableofcontents
\newpage
\section{Introduction}
\label{sIntr}

It is well-known that a single particle in $d+\ad$ dimensions gives
rise to towers of Kaluza-Klein modes in $d$ dimensions via
dimensional compactification
\cite{Kaluza:1921tu,Klein:1926tv,Duff:1986hr}.  The Kaluza-Klein
compactification works when $R^{d+\ad}$ is replaced by $R^d \times
M^\ad$ with some compact manifold $M^\ad$. Then, the expansion of a
solution of, say, Klein-Gordon equation in $R^d \times M^\ad$ in the
form $\phi(x,y) = \sum_n \phi_n(x)\psi_n(y)$ where $\psi_n (y)$ are
eigenfunctions of the Laplace operator on $M^\ad$ gives a tower of
Kaluza-Klein modes that starts with zero modes $\psi_{n_0} (x)$. The
Kaluza-Klein mass gap scales in the inverse size of the
compact manifold $M^\ad$. The mass spectrum is determined by the
spectrum of the Laplace operator on $M^\ad$.

For the text book $\ad=1$ example of a circle
of radius $R$, \ie $M^1=S^1$,%
\be\nn%
\psi_n(y) = e^{i\frac{ny}{R}},\quad n\in\mathbb{Z},\:
y\sim y+2\pi R\,.%
\ee%
The spectrum of $d$-dimensional particles resulting from a particle
of mass $m$ in $d+1$ dimensions is given by the well-known
formula%
\be\nn%
m_n^2=m^2+\frac{n^2}{R^2}.%
\ee%
{}From here it follows in particular that the spectrum becomes
continuous in the decompactifying limit $R\to \infty$,
that makes it difficult to  interpret
a particle in $R^{d+\ad}$ in terms of particles in $R^d$.

If one starts with the anti-de Sitter space $AdS_{d+\ad}$ instead of
$R^{d+\ad}$, the situation changes in two respects. The bad news is
that it is not clear what should be an $AdS$ analog of $R^d\times
M^\ad$ since $AdS_{d+\ad}$ is not a product of two manifolds. The
good news is however that the spectrum of states in $AdS_{d+\ad}$ is
discrete. As a result, a module of the $AdS_{d+\ad}$ algebra
$o(d+\ad-1,2)$ that describes states of one or another particle in
$AdS_{d+\ad}$ reduces to a countable number of irreducible modules
of $o(d-1,2)\oplus o(\ad)\subset o(d+\ad-1,2)$ that describe
particles in $AdS_{d}$ which form some $o(\ad)$ multiplets. The
spectrum remains discrete with the mass gap scale given by the
inverse $AdS$ radius $\lambda=\rho^{-1}$. That this should happen is
fairly clear from the structure of unitary lowest weight
$o(d,2)$-modules (see, e.g.,
\cite{Nicolai:1984hb,deWit:1999ui,Vasiliev:2004cm}). Since the
phenomenon is neither reduction (all degrees of freedom are kept
intact) nor compactification (nothing is compactified) we call it
dimensional degression. Note that our approach differs from that of
\cite{Biswas:2002nk} (see also \cite{Francia:2008hd}) where the
radial reduction of the flat to curved space was carried out.

The main aim of this paper is to check how the dimensional
degression occurs in terms of  fields. We analyze the problem in
full generality for the example of a scalar field in $AdS_{d+\ad}$
showing that the spectrum of scalar field modes indeed matches the
pattern of the dimensionally reduced unitary modules. We also
demonstrate that our results reproduce those obtained previously by
Metsaev \cite{Metsaev:2000qb} by a different method based on a
specific Ansatz. In addition, our approach reproduces the shadow
unitary states that exist for sufficiently low energy eigenvalues
with negative $m^2$.

The obtained results may have several applications. First of all,
this is the first step towards the analysis of the dimensional
compactification-like effects in higher-spin (HS) theories in
$AdS_d$. This analysis may be useful, in particular, for
understanding a mechanism of spontaneous breakdown of HS gauge
symmetries. Indeed, we show that the dimensional degression gives
rise to the gauge invariant (i.e., Stueckelberg) formulation of
massive fields. As such it is analogous to the standard torus
compactification mechanism used to derive massive HS theories from
the massless ones in one higher dimension \cite{SS,RSS,ADY}. In
other words, the degression mechanism in $AdS$ may help to uncover
the structure of the Higgs sector in HS gauge theory. In particular,
it would be interesting to see how the gauge invariant formulation
\cite{Zinoviev:2001dt,Buchbinder:2006ge,
Metsaev:2006zy,Buchbinder:2006nu} of massive HS fields
\cite{Singh:1974qz,
Singh:1974rc,Metsaev:2003cu,Buchbinder:2005ua,Francia:2007ee} in
$AdS_d$ results from dimensional degression.

The layout of the rest of the paper is as follows: in Section
\ref{sUnitRepr} relevant facts on the unitary $o(d,2)$-modules are
recalled and the analysis of the branching of the unitary
$o(d,2)$-modules into $o(d-1,2)$-modules is performed (detailed
proofs are given in Appendix). In Section
\ref{sRedOneDim} the field-theoretical analysis of the dimensional
degression of $AdS_{d+1}$ to $AdS_{d}$ is done  for the case of a
spin zero field of arbitrary mass and massless fields of spins
one-half, one, and two. The generalization to several extra
coordinates is given in Section \ref{sRedMoreDim} for a
scalar field. Conclusions and perspectives are discussed in
Conclusion.

\section{Group-theoretical analysis}
\label{sUnitRepr}

 The generators $T^{AB}=-T^{BA}$ ($A,B =-1, 0, \ldots , d$) of
 the symmetry algebra $o(d,2)$ of $AdS_{d+1}$ have the commutation
 relations
\be\nn%
\left[T^{AB}\,,T^{CD}\right]= \eta^{BC} T^{AD} - \eta^{AC} T^{BD}
-\eta^{BD}
T^{AC} +\eta^{AD} T^{BC}\,,%
\ee%
where  $\eta^{AB}$ is the invariant symmetric  form of $o(d,2)$. We use the mostly minus convention with
$\eta^{-1-1} = \eta^{00}=1$ and $\eta^{ab} = -\delta^{ab}$ for the
space-like values of $A=a=1\ldots d$. The $AdS_{d+1}$ energy
operator is%
\be\nn%
E = i T^{-1\,0}\,.%
\ee%
The noncompact generators of $o(d,2)$ are%
\be\nn%
T^{\pm a} =iT^{0a}\mp  T^{-1\,a}\,,%
\ee%
\be\nn%
[E, T^{\pm a}] = \pm T^{\pm a}\,,\qquad%
[T^{- a} , T^{+ b} ] = 2(\delta^{ab} E + T^{ab} )\,.%
\ee%
The compact generators $T^{ab}$ of $o(d)$ commute with $E$. The
generators $T^{AB}$ are anti-Hermitian, $(T^{AB})^\dagger = -
T^{AB}$. Hence,
\be\nn%
E^\dagger = E\,,\qquad (T^{\pm a})^\dagger = T^{\mp a}\,,\qquad
(T^{ab})^\dagger = - T^{ab}\,.%
\ee%

An irreducible bounded energy unitary $o(d,2)$-module
$\mathcal{H}(E_0,{\bf s})_{o(d,2)}$  is characterized by the
eigenvalue $E_0$ of $E$ and the weight ${\bf s}=(s_1,s_2,\ldots,
s_{[d/2]})$ of $o(d)$ which refer to the lowest energy (vacuum)
states $\left|E_0,{\bf s}\right>$ of $\mathcal{H}(E_0,{\bf
s})_{o(d,2)}$ that satisfy $T^{-a} \left|E_0,s\right>=0$ and form a
finite dimensional module of $o(d)\oplus o(2) \subset o(d,2)$. $E_0$
characterizes the energy of the ground state of a field (related to
its mass) and ${\bf s}$ characterizes the spin of a field.

Recall that in the complex case of $o(d|\mathbb{C})$, all $s_i$ are
integers in the bosonic case and half-integers in the fermionic case
that are all non-negative except $s_{\frac{d}{2}}$ for even $d$,
which may be positive, negative or zero. Two sign possibilities for
$s_{\frac{d}{2}}$ correspond to selfdual and anti-selfdual
representations $o(d|\mathbb{C})$ (equivalently, left and right
chiral spinor representations in the fermionic case). The structure
of irreducible representations of the real algebra $o(d|\mathbb{R})$
is more subtle because the (anti)selfduality or chirality conditions
may or may not be compatible with the reality (Majorana) conditions.
We will still use the same spin notations as in the complex case,
assuming that the chirality or self-duality conditions are imposed
on the doubled set of tensors. The resulting module may be
irreducible if the chirality or self-duality conditions are
incompatible with the reality condition or double-reducible
otherwise. In practice, this convention is not essential for the
analysis of bosonic fields in this paper because we mostly consider
the case with $s_{\frac{d}{2}}=0$. In the fermionic case, however,
it corresponds to the consideration of the left or right spinors,
neglecting the $d$-dependent analysis of Majorana conditions. Such a
convention just matches the field theoretical analysis of this paper
that will be performed in terms of Dirac spinors.

The requirement of unitarity results in the constraints on the
energy $E_0$ and the spin ${\bf s}$. For scalar and spinor
they are \cite{Ferrara:1997dh,Laoues:1998ik,Ferrara:2000nu}
\begin{equation}\label{UnitCondLowest} %
E_0 \geq E_0({\bf 0}) = \frac{d-2}{2}\quad\text{if}\ s=0\:,\qquad %
E_0 \geq E_0({\bf\scriptstyle \frac 1 2}) = \frac{d-1}{2}\quad\text{if}\ s=\frac 1 2 \:.%
\end{equation} As shown by Metsaev
\cite{Metsaev:1995re,Metsaev:1998}, for a generalized spin
${\bf s}$   with $s_1=\ldots =s_p>s_{p+1}\geq\ldots\geq
s_{[d/2]}$ and $s_1\geq 1$,  the  unitarity constraint is
\be\label{UnitCondS} %
E_0 \geq E_0({\bf s}) = s_p - p - 1 + d\: .%
\ee%

Let $\Verma{E_0}{s}{d}$ be a generalized Verma module induced from
some irreducible $o(d)\oplus o(2)$ vacuum module $\left|E_0,{\bf
s}\right>$ annihilated by $T^{-a}$. It is spanned by the states
\be\label{states}%
T^{+a_1}\ldots T^{+a_M} \left|E_0,{\bf s}\right>,%
\ee%
with various $M$. For the unitary case $E_0>E_0({\bf s})$,
$\Verma{E_0}{{\bf s}}{d}=\mathcal{H}(E_0,{\bf s})_{o(d,2)}$. At the
boundary of the unitarity region $E_0=E_0({\bf s})$,
$\Verma{E_0}{s}{d}$ contains a singular submodule $S$ of null
states. This has to be factored out to obtain a unitary
$o(d,2)$-module $\mathcal{H}(E_0,s)_{o(d,2)}=\Verma{E_0}{s}{d}/S$.

In this paper, we consider only the case of totally symmetric
representations with ${\bf s}=(s,0,\ldots,0)$ or ${\bf
s}^\pm=(s,\frac 1 2,\ldots,\pm\frac 1 2)$ if $d$ is even and ${\bf
s}=(s,\frac 1 2,\ldots,\frac 1 2)$ if $d$ is odd. The respective
unitary modules are denoted by $\mathcal{H}(E_0,{s})_{o(d,2)}$.

The branching rules for the unitary lowest weight modules associated
with fields in $AdS_{d+1}$ are summarized by the following Theorems.

\begin{Thr}%
\label{Thr_bos}%
For an integer spin $s$, the $o(d,2)$-module $\Verma{E_0}{s}{d}$ with
$E_0\geq E_0({\bf s})$ has the following branching into
$o(d-1,2)$--modules
\be
\label{dec}
\Verma{E_0}{s}{d}\simeq \sum^s_{q=0}\sum^\infty_{k=0}\oplus \Verma{E_0+k}{q}{d-1} .%
\ee%
\end{Thr}%
\begin{Thr}%
\label{Thr_fer}%
For a half-integer spin $s$ and $E_0\geq E_0({\bf s})$,
the $o(d,2)$-module $\Verma{E_0}{s^\pm}{d}$ for even $d$ and
$\Verma{E_0}{s}{d}$ for odd $d$  have, respectively, the following
branchings into $o(d-1,2)$--modules
\be \label{dec_even}
\Verma{E_0}{s^\pm}{d}\simeq \sum^{s}_{q=\frac 1 2}\sum^\infty_{k=0}\oplus \Verma{E_0+k}{q}{d-1},%
\ee%
\be \label{dec_odd}%
\Verma{E_0}{s}{d}\simeq \sum^{s^+}_{q^+=\frac 1
2}\sum^\infty_{k=0}\oplus \Verma{E_0+k}{q^+}{d-1}
\oplus \sum^{s^-}_{q^-=\frac 1 2}\sum^\infty_{k=0}\oplus \Verma{E_0+k}{q^-}{d-1}.%
\ee%
\end{Thr}%
For the proofs see Appendix.


At the boundary energy $\Verma{E_0(s)}{s}{d}$ contains a singular
submodule $S$ which  is
\be\nn%
S_{(s)}\simeq \Verma{d+s-1}{s-1}{d}%
\ee%
for $s\geq 1$ and
\be\nn %
S_{(0)}\simeq \Verma{\frac{d+2}{2}}{0}{d}\:,\quad S_{(\frac 1
2)}\simeq \Verma{\frac{d+1}{2}}{\frac 1 2}{d}\:
\ee%
for $s=0$ or $1/2$. It is important that $S$ itself can be endowed
with an $o(d,2)$ invariant positive-definite form (which, however,
cannot be extended to an invariant form on $\Verma{E_0(s)}{s}{d}$),
hence forming a unitary $o(d,2)$-module. This allows us to apply
Theorems \ref{Thr_bos} and \ref{Thr_fer} to its branching as well.
For integer spins we obtain
\begin{align}\label{S}%
S_{(s)} &\simeq \sum^{s-1}_{q=0}\sum^\infty_{k=1}\oplus \Verma{k+d+s-2}{q}{d-1}\:,\\ %
S_{(0)} &\simeq \sum^\infty_{k=2}\oplus \Verma{\frac{d-2}{2}+k}{0}{d-1} \label{S_0}%
\end{align} %
for any $d$. For half-integer spins
\begin{align}%
S_{(s)} &\simeq \sum^{s-1}_{q=\frac 1 2}\sum^\infty_{k=1}\oplus \Verma{k+d+s-2}{q}{d-1}\:,\\ %
S_{(\frac 1 2)} &\simeq \sum^\infty_{k=1}\oplus \Verma{k+\frac{d-1}{2}}{\frac 1 2}{d-1} %
\end{align} %
for even $d$ and
\begin{align} %
S_{(s)} &\simeq \sum^{s^+-1}_{q^+=\frac 1 2}\sum^\infty_{k=1}\oplus \Verma{k+d+s-2}{q^+}{d-1}%
\oplus\sum^{s^--1}_{q^-=\frac 1 2}\sum^\infty_{k=1}\oplus
\Verma{k+d+s-2}{q^-}{d-1}\:,\\%
S_{(\frac 1 2)} &\simeq \sum^\infty_{k=1}\oplus \Verma{k+\frac{d-1}{2}}{{\frac 1 2}^+}{d-1}%
\oplus\sum^\infty_{k=1}\oplus
\Verma{k+\frac{d-1}{2}}{{\frac 1 2}^-}{d-1}%
\end{align} %
for odd $d$.

This gives
\begin{Cor}\label{ThSpin}%
The unitary $o(d,2)$-module ${\cal H}(d+s-2,s)_{o(d,2)}$ with
integer spin $s=1,2,\ldots$ has the following branching into
$o(d-1,2)$ unitary modules
\be\nn%
{\cal H}(d+s-2,s)_{o(d,2)}\simeq \sum^\infty_{k=0}\oplus
\Verma{k+d+s-2}{s}{d-1}\oplus
\sum^{s-1}_{q=0}\oplus \Verma{d+s-2}{q}{d-1}.%
\ee%
\end{Cor}

\begin{Cor}\label{ThSpin_fer}
The unitary  $o(d,2)$-module ${\cal H}(d+s-2,s^\pm)_{o(d,2)}$ for
even $d$ and ${\cal H}(d+s-2,s)_{o(d,2)}$ for odd $d$ with
half-integer spin $s=\frac 3 2,\frac 5 2,\ldots$ have the following
branching into the $o(d-1,2)$ unitary modules
\be\nn%
{\cal H}(d+s-2,s^\pm)_{o(d,2)}\simeq \sum^\infty_{k=0}\oplus
\Verma{k+d+s-2}{s}{d-1}\oplus \sum^{s-1}_{q=\frac 1 2}\oplus
\Verma{d+s-2}{q}{d-1}\: ,%
\ee%
\begin{align*}%
{\cal H}(d+s-2,s)_{o(d,2)}\simeq \sum^\infty_{k=0}\oplus
\Verma{k+d+s-2}{s^+}{d-1}\oplus\sum^\infty_{k=0}\oplus \Verma{k+d+s-2}{s^-}{d-1}\\%
\oplus \sum^{s^+-1}_{q^+=\frac 1 2}\oplus
\Verma{d+s-2}{q^+}{d-1}\oplus \sum^{s^--1}_{q^-=\frac 1 2}\oplus
\Verma{d+s-2}{q^-}{d-1} \: .
\end{align*} %
\end{Cor}

\begin{Cor}\label{ThSpinLess}
$o(d,2)$ singleton modules have the following branchings into
$o(d-1,2)$ unitary modules
\be\nn%
{\cal H}\left(\frac{d-2}{2},0\right)_{o(d,2)}\simeq
\Verma{\frac{d-2}{2}}{0}{d-1}\oplus \Verma{\frac{d}{2}}{0}{d-1}%
\ee%
for any $d$,
\be\nn%
{\cal H}\left(\frac{d-1}{2},{\frac 1 2}^\pm\right)_{o(d,2)}\simeq
\Verma{\frac{d-1}{2}}{\frac 1 2}{d-1}%
\ee%
for even $d$ and
\be\nn%
{\cal H}\left(\frac{d-1}{2},\frac 1 2\right)_{o(d,2)}\simeq
\Verma{\frac{d-1}{2}}{{\frac 1 2}^+}{d-1}\oplus \Verma{\frac{d-1}{2}}{{\frac 1 2}^-}{d-1}%
\ee%
for odd $d$.
\end{Cor}

These results have the following field-theoretical interpretation.
The dimensional degression of one dimension for the spin $s$ field
with the energy $E_0$ above the unitarity bound should
give rise to the set of fields with all spins $0,1,\ldots,s$ for the
bosonic case and $\frac 1 2,\frac 3 2,\ldots, s$ for the fermionic
one. In the latter case each spin is doubled if the dimension of
$AdS_{d+1}$ is even. The energy spectrum is $E_0+n$, $n=0,1,\ldots$
for every spin. Note that the dimensional degression of a massless
scalar or Dirac field in $AdS_{d+1}$ does not produce a massless
field in $AdS_d$. Similarly, a spin $s\geq 1$ massless field in
$d+1$ dimensions has the lowest energy $E_0=d+s-2$. Its degression
by one dimension gives a set of spin $s$ fields with energies $E_n =
n+d+s-2$, $n=0,1,\ldots$ plus a set of fields with spins
$0,1,\ldots, s-1$ or $\frac 1 2,\frac 3 2,\ldots, s-1$ and the
energy $d+s-2$. In the fermion case each field is doubled if the
dimension of $AdS_{d+1}$ is even. All of these fields are massive in
$d$ dimensions. Dimensional degression of a scalar singleton gives
 massless scalar fields.  The dimensional degression
of a fermion singleton yields one massless Dirac field if the
original space is odd-dimensional or two otherwise. These results
agree with the well-known fact that singletons are conformal
fields in one lower dimension \cite{Dirac:1963ta, Flato:1978qz,
Flato:1980zk,Laoues:1998ik,Vasiliev:2004cm}.

The obtained results make it easy to analyze the dimensional
degression over several dimensions. Let us consider the example of
the scalar representation. The branching of an $o(d+\ad-1,2)$-module
$\Verma{E_0}{0}{d+\ad-1}$ as $o(d-1,2)$-modules yields
\be\label{UnitMore} %
\Verma{E_0}{0}{d+\ad-1}\simeq
\sum^\infty_{k=0}\oplus \binom{k+\ad-1}{\ad-1} \Verma{E_0+k}{0}{d-1}.%
\ee%
Thus, the dimensional degression of a scalar field in $AdS_{d+\ad}$
over $\ad$ coordinates gives the set of scalar fields of the
energies $E_0+k$ and the multiplicities $\binom{k+\ad-1}{\ad-1}$,
where $k=0,1,\ldots$ The degeneracy manifests that the scalar fields
of energy $E_0+k$ carry the module
\be\nn%
\Verma{E_0+k}{0}{d-1}\otimes\bigoplus\limits^{\left[\frac{k}{2}\right]}_{l=0}D(k-2l)%
\ee%
of $o(d-1,2)\oplus o(\ad)\subset o(d+\ad-1,2)$, where $D(k-2l)$ is
an $o(\ad)$-module of weight $(k-2l,0,\ldots)$. The binomial
coefficients in (\ref{UnitMore}) are just the dimensions of
$\bigoplus\limits^{\left[\frac{k}{2}\right]}_{l=0}D(k-2l)$
$o(\ad)$-modules.

\section{Field-theoretical degression of one dimension}
\label{sRedOneDim}

The aim of this section is to show how the results of the
group-theoretical analysis of Section \ref{sUnitRepr} are reproduced
in the field-theoretical models.

Let us describe a $(d+1)$-dimensional anti-de Sitter space as a hyperboloid
\be\nn%
y_{-1}^2+y_0^2-y_1^2-\ldots-y_d^2=1%
\ee%
in $R^{d+2}$. We choose the following coordinates in $AdS_{d+1}$
\be\nn%
y_{-1}=\cosh(z)\tilde{y}_{-1},\ y_{0}=\cosh(z)\tilde{y}_{0},\
\ldots,\  y_{d-1}
=\cosh(z)\tilde{y}_{d-1},\  y_d=-\sinh(z),%
\ee%
where $z \in (-\infty,+\infty)$ and
$\tilde{y}_{-1},\ldots,\tilde{y}_{d-1}$ satisfy
\be\nn%
\tilde{y}_{-1}^2+\tilde{y}_0^2-\tilde{y}_1^2-\ldots-\tilde{y}_{d-1}^2=1,
\ee%
\ie describe $AdS_d$. The line element of $AdS_{d+1}$ has the form
\be\label{MetricDecomposition}%
ds^2_{AdS_{d+1}}=\cosh^2(z)ds^2_{AdS_d}-dz^2,%
\ee%
where $ds^2_{AdS_d}$ is the line element of $AdS_d$. The warp factor
$\cosh^2(z)$ manifests that $AdS_{d+1}$ is not a direct product of
$AdS_d$ with some other manifold.

We use the following index conventions
\be\label{IndexDecomposition}%
\alpha,\beta\ldots  = 0,1,\ldots,d\,,\quad
\mu,\nu\ldots  = 0,1,\ldots,d-1\,,\quad \bullet \equiv d\:, %
\ee%
\ie lower case Greek letters $\alpha,\beta\ldots$ correspond to
coordinates of $AdS_{d+1}$, lower case Greek letters $\mu,\nu\ldots$
correspond to coordinates of $AdS_{d}$, and $\bullet$ denotes the
extra coordinate of $AdS_{d+1}$ compared to $AdS_d$.

Let us comment on the relation between mass and energy in $AdS_d$.
We refer to the term $m^2$ in the equation
\be\label{eq_ex} %
(\Box_{AdS_d}+m^2)\psi_{\mu_1\ldots\mu_s}(x)=0\:, %
\ee%
where $\Box_{AdS_d}\equiv D^\mu D_\mu$ is the Laplace-Beltrami operator,
and $D_\mu$ is the $AdS_d$ covariant derivative, as the mass square of
the symmetric bosonic field $\psi_{\mu_1\ldots\mu_s}(x)$. The
relation with the lowest energy is \cite{Metsaev:1997nj}
\be\label{MassEnergy} %
m^2=E(E-d+1)-s\:.%
\ee%
Equation (\ref{MassEnergy}) yields two solutions for the energies
for the same mass. In the case of spin $1,2,\ldots$, only
the highest energy
obeys the unitary condition (\ref{UnitCondS}). In the scalar field
case there are two options. If $m^2 > -(d-3)(d+1)/4$, only the
highest energy satisfies the unitary condition
(\ref{UnitCondLowest}). If $-(d-1)^2/4\leq m^2 \leq -(d-3)(d+1)/4$,
both solutions for the energy obey the unitarity condition. Thus,
in the latter case, the scalar field equation (\ref{eq_ex}) admits
two types of normalizable solutions. Note
that they are not mutually orthogonal with respect to the standard
norm \cite{Breitenlohner:1982jf,Balasubramanian:1998sn} (see also
Subsection \ref{scal}).

\subsection{Scalar field}
\label{scal}
The free wave equation of a massive scalar field in $AdS_{d+1}$ is
\be\label{RedOneDimAdSEq}%
(\Box_{AdS_{d+1}}+M^2)\Phi(x,z)=0,%
\ee%
where $x$ are local coordinates of $AdS_d$. In the coordinates
(\ref{MetricDecomposition}) the Laplace-Beltrami operator
$\Box_{AdS_{d+1}} = \frac{1}{\sqrt{|g|}}\pd_\alpha
g^{\alpha\beta}\sqrt{|g|}\pd_\beta$ is
\be\label{RedOneDimBox}%
\Box_{AdS_{d+1}} = \frac{1}{\cosh^2(z)}\Box_{AdS_d} - %
\frac{1}{\cosh^d(z)}\pd_z\cosh^d(z)\pd_z,
\ee%
where $\pd_z\equiv\frac{\pd}{\pd z}$.

Recall, that the space of solutions is endowed with the inner
product defined as the electric charge
\begin{align}
\label{FieldTheorNorm}
(\phi,\psi)\equiv(\phi^+,\psi^+)=
i\int_\Sigma d\sigma^\alpha\phi^{+*}\overleftrightarrow{\pd_\alpha}\psi^+%
=i\int_{t=const}g^{00}\sqrt{|g_{AdS_{d+1}}|}d^dx\;\phi^{+*}
\overleftrightarrow{\pd_0}\psi^+%
\end{align}
of the positive frequency parts of two real solutions $\phi$ and $\psi$
of the wave equation. $\Sigma$ is a space-like surface that can be
chosen to be a surface of constant time $t$. Since the electric
charge conserves, so defined inner product is formally independent
of the integration surface choice. The conditions of finiteness and
actual conservation for the norm impose certain boundary conditions
that restrict the energies of solutions.

Let us look for solutions of the equation (\ref{RedOneDimAdSEq}) in the
form
\be\label{RedOneDimFourier} %
\Phi(x,z) = \sum \limits^\infty_{N=0} \phi_N(x)P_N(z),
\ee%
where $P_N (z)$ form an orthogonal complete set with respect to
the norm
\be \label{RedOneDimNorm}%
\int\limits^\infty_{-\infty}dz\cosh^{d-2}(z)P_N(z)P_M(z) = \delta_{NM}%
\ee%
inherited from the norm (\ref{FieldTheorNorm}) in the $z$ sector.
The conservation of the norm (\ref{FieldTheorNorm}) requires
solutions to satisfy the boundary conditions
\be\label{Boundary_Scalar} %
\left.\cosh^d(z)\left(P_N(z)\pd_z P_M(z) - P_M(z)\pd_z P_N(z)\right)
\right|_{z\rightarrow\pm\infty} = 0. %
\ee%

Substitution of the expansion (\ref{RedOneDimFourier}) into
Eq.~(\ref{RedOneDimAdSEq}) yields a tower of massive scalar
fields in $AdS_d$
\be\label{RedOneDimScalarEq} %
(\Box_{AdS_d}+m_N^2)\phi_N(x)=0 %
\ee%
along with the following equation on $P_N(z)$
\be\label{RedOneDimPEq}%
\pd_z^2 P_N(z)+d\tanh(z)\pd_zP_N(z)+\left[\frac{m_N^2}{\cosh^2(z)}-M^2\right]P_N(z)=0.%
\ee%

By the change of variables $\tan(\varphi) = \sinh(z)$ and
$p_N(\varphi)=\cos^{-\frac{d-1}{2}}(\varphi)P_N(\varphi)$ with the
coordinate $\varphi\in \left(-\frac\pi 2,\frac\pi 2\right)$ the
equation (\ref{RedOneDimPEq}) and norm (\ref{RedOneDimNorm}) are
mapped to the one-dimensional quantum mechanics on the interval
$\left(-\frac\pi 2,\frac\pi 2\right)$ with the energy
$E=m^2_N+\frac{(d-1)^2}{4}$ and the P\"{o}schl-Teller potential
\cite{Flugge}
$U(\varphi)=\cos^{-2}(\varphi)\left(M^2+\frac{d^2-1}{4}\right)$
\be\label{Shred} %
\pd_\varphi^2\:p_N+%
\left[m^2_N+\frac{(d-1)^2}{4}-\cos^{-2}(\varphi)\left(M^2+\frac{d^2-1}{4}\right)\right]p_N=0\:,%
\ee%
\be %
\int\limits^{\frac\pi 2}_{-\frac\pi 2}d\varphi\:p_N(\varphi)p_M(\varphi)=\delta_{NM}\:.%
\ee%

The orthogonality  and boundary conditions (\ref{RedOneDimNorm}) and
(\ref{Boundary_Scalar}) constrain solutions to the ``irregular''
class
\begin{align}%
&P^{irr}_N(z)=N^{-\kappa}_N\cosh^{-\frac{d-2\kappa}{2}}{(z)}{}_2F_1\left(N-2\kappa+1,-N;1-\kappa;\frac{1+\tanh
(z)}{2}\right),\label{ScalarPirr} \\
&M^2\in \left[-\frac{d^2}{4},-\frac{d^2}{4}+1\right)\nn
\end{align}%
and ``regular'' class
\begin{align}%
&P^{reg}_N(z)=N^\kappa_N\cosh^{-\frac{d+2\kappa}{2}}{(z)}%
{}_2F_1\left(N+2\kappa+1,-N;1+\kappa;\frac{1+\tanh(z)}{2}\right),\label{ScalarPreg} \\%
&M^2\in \left[-\frac{d^2}{4},+\infty\right)\:,\nn
\end{align}%
where ${}_2F_1(a,b;c;x)$ is the hypergeometric function
(see e.g. \cite{GrandshteynRyzhik}),
\be\nn%
\kappa\equiv\sqrt{\frac{d^2}{4}+M^2}%
\ee%
and the normalization factor%
\be\label{NormFactor} %
N^\kappa_n=\frac{\sqrt{2\kappa+2n+1}}{2^{\kappa+\frac 1
2}\Gamma(\kappa+1)}\sqrt{\frac{\Gamma(2\kappa+n+1)}{n!}}
\ee%
is fixed by (\ref{RedOneDimNorm}).

For the ``irregular'' solutions the lowest energies and the related
mass spectrum (\ref{MassEnergy}) are given by the formulas
\begin{align}\label{RedOneDimEnergySing}%
E_N &= N+\frac{d-2\kappa}{2}\:,\\%
m^2_N &=\left(N-\kappa+\frac 1 2\right)^2-\frac{(d-1)^2}{4}\:.\nn%
\end{align}%
The ``regular'' solutions have the energies%
\be\label{RedOneDimEnergyReg}%
E_N = N+\frac{d+2\kappa}{2}
\ee%
and the masses%
\be\nn%
m^2_N =\left(N+\kappa+\frac 1 2\right)^2-\frac{(d-1)^2}{4}\:.%
\ee%

Let us discuss the normalization conditions for ``regular'' and
``irregular'' solutions in more detail. First of all, we observe
that the integral (\ref{RedOneDimNorm}) absolutely converges in the
union of ``regular'' and ``irregular'' spaces. So, the key role is
played by the boundary conditions (\ref{Boundary_Scalar}). As is
obvious from (\ref{ScalarPreg}), the regular solutions decrease as
$\exp\left(-\frac{d+2\kappa}{2}|z|\right)$ at $z\to \pm \infty$
($N=0,1,\ldots$) so that each of the terms in
(\ref{Boundary_Scalar}) vanishes. For the ``irregular'' class
(\ref{ScalarPirr}) the leading contribution to each term in
(\ref{Boundary_Scalar}) increases at infinity as $\exp(2\kappa
|z|)$, but  it cancels between the two terms both at $z\to +\infty$
and at $z\to -\infty$. The subleading terms behave as
$\exp(2(\kappa-1)|z|)$ and vanish for $\kappa <1$, which is just the
domain where the ``irregular" solutions form a unitary module. If a
``regular'' solution is paired with an ``irregular'' one, the
condition (\ref{Boundary_Scalar}) is not satisfied because different
terms in (\ref{Boundary_Scalar}) tend to different constants that
cannot cancel out. As a result, the $o(d,2)$-invariant norm
(\ref{FieldTheorNorm}) is well-defined for the spaces of ``regular''
and ``irregular'' solutions separately, but not for their union.
This implies in particular that the ``regular'' and ``irregular''
solutions do not possess definite mutual orthogonality properties.

Using that both $P_N^{irr}$ and $P_N^{reg}$ form  orthonormal bases,
we can perform the  dimensional degression at the action level for
each class. Indeed, plugging (\ref{RedOneDimBox}) and
(\ref{RedOneDimFourier}) into the scalar field action in $AdS_{d+1}$
\be\nn%
S^{AdS_{d+1}}=\frac 1 2\int\Vol{d+1}\Phi(\Box_{AdS_{d+1}}+M^2)\Phi\:, %
\ee%
we observe that
\be\nn%
S^{AdS_{d+1}}=\sum\limits^\infty_{N=0}S_N^{AdS_d},%
\ee%
where
\be\nn %
S_N^{AdS_d}=\frac 1 2\int
\Vol{d}\phi_N(\Box_{AdS_d}+m^2_N)\phi_N%
\ee%
is the action of a scalar field of the mass $m^2_N$ in $AdS_d$. Here
$\Vol{d}\equiv \sqrt{|g_{AdS_d}|}d^dx$ is the
$AdS_d$--invariant volume element.

These results are consistent with the group-theoretical analysis of
Section \ref{sUnitRepr}. In accordance with Theorem \ref{Thr_bos},
for a fixed energy $E_0$, the dimensional degression gives the set
of scalar fields with the energy spectrum $E_0+k$, $k=0,1,\ldots$
The appearance of  two types of solutions, namely, ``regular'' and
``irregular'' ones, results from the ambiguity in the relation
between the square mass and the lowest energy in (\ref{MassEnergy}).
Note that the dimensional degression of a massless scalar field over
one coordinate produces no massless fields.

Let us now compare our results with those of Metsaev
\cite{Metsaev:2000qb} who analyzed  analogous problem for the
conformal mass case $m^2=-\frac{d^2-1}{4}$ where the free wave
equation is conformal invariant. In \cite{Metsaev:2000qb}, the
following Fourier expansion was used
\be\label{MetsaevFourie}%
\Phi(x,X,Z) = \frac{1}{\sqrt{z}} \sum\limits_{n\in\mathbb{Z}}
e^{in\varphi}\phi_n(x,z),%
\ee%
where $x,X$ and $Z$ are Poincar\'e coordinates of $AdS_{d+1}$
\be\nn%
ds^2_{AdS_{d+1}}=\frac{1}{Z^2}(dx^adx_a+dX^2+dZ^2)\:,\quad Z>0\: %
\ee%
(in the rest  of this section we use the mostly plus metric
convention of \cite{Metsaev:2000qb}) and the coordinates
$z$ and $\varphi$ are defined by
\be\nn%
Z=z\sin\varphi,\quad X=z\cos\varphi,%
\ee%
with $\varphi\in(0,\pi)$ and $z>0$. Plugging the Fourier expansion
into the scalar field equation gives the mass spectrum
\be\nn%
m^2_n=n^2-\frac{(d-1)^2}{4}%
\ee%
and the energy spectrum%
\be\nn%
E_n=n+\frac{d-1}{2}\:, \quad n=0,1,\ldots%
\ee%

To establish the precise correspondence with our results, we should
however use the Fourier expansion
\be\label{twobranch} %
\Phi(x,X,Z) = \frac{1}{\sqrt{z}}
\sum\limits^\infty_{n=0}\psi_n\cos(n\varphi)\quad %
\ \text{or}\quad %
\Phi(x,X,Z) = \frac{1}{\sqrt{z}}
\sum\limits^\infty_{n=1}\xi_n\sin(n\varphi)\:.%
\ee%
This yields the energy spectra
\be\nn%
E_n^\psi=n+\frac{d-1}{2}\:,\quad E_{n+1}^\xi=n+\frac{d+1}{2}\:,\quad n=0,1,\ldots%
\ee%
The relationship between $\psi_n,\:\xi_n$ and $\phi_n$ is
\[%
\phi_n=\left\{
    \begin{array}{ll}
        \frac{\psi_n-i\xi_n}{2}\quad n=1,2,\ldots,\\
        \psi_0\quad n=0,\\
        \frac{\psi_n+i\xi_n}{2}\quad n=-1,-2,\ldots.
    \end{array}
\right.\]%

We see that the expansion (\ref{MetsaevFourie}) of
\cite{Metsaev:2000qb} mixes the two branches (\ref{twobranch}).
Note that the states described by (\ref{MetsaevFourie}) do
not form an orthogonal set with respect to the invariant norm
(\ref{FieldTheorNorm}) because the classes of ``regular'' and
``irregular'' solutions do not possess definite orthogonality
properties with respect to the norm (\ref{FieldTheorNorm}).

\subsection{Massless Dirac field}%

The action of Dirac spinor field $\widetilde{\Psi}$ of mass $M$ in
$AdS_{d+1}$ is
\be%
S = \int\Vol{d+1} \overline{\widetilde{\Psi}}\left(i\Gamma^\alpha
E^{\underline{\alpha}}_\alpha\nabla_{\underline{\alpha}}-M\right)\widetilde{\Psi}\:,
\ee%
where
\be%
\{\Gamma^\alpha,\Gamma^\beta\}=2\eta^{\alpha\beta}\:, \quad \eta^{\alpha\beta}=(+,-,\ldots,-)\:.%
\ee%
The covariant derivative is
\be%
\nabla_{\underline{\alpha}}=\pd_{\underline{\alpha}}-\frac i 2
\Omega^{\alpha\beta}_{\underline{\alpha}}M_{\alpha\beta}\:,\quad %
M_{\alpha\beta}=\frac i 4[\Gamma_\alpha,\Gamma_\beta]\:. %
\ee%
$E^{\underline{\alpha}}_\alpha$ and
$\Omega^{\alpha\beta}_{\underline{\alpha}}$ are inverse vielbein and
spin-connection of $AdS_{d+1}$.

For the sake of simplicity let us consider only the massless case of
$M=0$. Let us note that for the even-dimensional $AdS_{d+1}$ a Weyl
spinor field does not form a  $o(d,2)$--module\footnote{This is because
 the spinor representation of $o(d,2)$ beside
generators $T^{\alpha\beta}=\frac{1}{4}[\gamma^\alpha,\gamma^\beta]$
corresponding its Lorentz subalgebra $o(d,1)$ involves
$T^{\alpha\:-1}=\frac{i}{2}\gamma^\alpha$ that do not commute with
the Weyl projectors.}. In this case it is convenient to
choose the following coordinates
\be\label{MetricDirac} %
ds^2_{AdS_{d+1}} =
\cos^{-2}(\varphi)\left(ds^2_{AdS_d}-d\varphi^2\right)\:,\quad \varphi\in\left(-\frac \pi 2,\frac \pi 2\right)%
\ee%
related to the coordinates (\ref{MetricDecomposition}) via
$\tan(\varphi)=\sinh(z)$. In these coordinates, the vielbein and the
spin-connection of $AdS_{d+1}$ can be expressed via the vielbein
$e^\mu_{\underline{\mu}}$ and the spin-connection
$\omega^{\mu\nu}_{\underline{\mu}}$ of $AdS_d$ as follows
\be%
\begin{array}{ll}
E^\mu_{\underline{\mu}}=\cos^{-1}(\varphi)e^\mu_{\underline{\mu}}\:,\quad
&E^{\bullet}_{\underline{\bullet}}=\cos^{-1}(\varphi)\:, \\%
E^\mu_{\underline{\bullet}}=0\:,
&E^{\bullet}_{\underline{\mu}}=0\:, \\%
\Omega^{\mu\nu}_{\underline{\mu}}=\omega^{\mu\nu}_{\underline{\mu}}\:,
&\Omega^{\mu\bullet}_{\underline{\mu}}= -\tan(\varphi)e^\mu_{\underline{\mu}}\:, \\%
\Omega^{\mu\nu}_{\underline{\bullet}}=0\:,
&\Omega^{\mu\bullet}_{\underline{\bullet}}=0\:.
\end{array}%
\ee%
Also substituting
\be%
\widetilde{\Psi}(x) = \cos^{\frac d 2}(\varphi)\Psi(x)\,,%
\ee%
the Dirac action in coordinates (\ref{MetricDirac}) takes the form
\be\label{Dirac_Action} %
S = \int \Vol{d} \int\limits^{\frac \pi 2}_{-\frac \pi 2} d\varphi\
\overline{\Psi}\left(i\Gamma^\mu e^{\underline{\mu}}_\mu
D_{\underline{\mu}}+ i\Gamma^\bullet\pd_\varphi\right)\Psi\:,
\ee%
where $D_{\underline{\mu}}=\pd_{\underline{\mu}}-\frac i 2
\omega^{\mu\nu}_{\underline{\mu}}M_{\mu\nu}$ is the covariant
derivative in $d$ dimensions.

The norm on the space of solutions is defined in the standard way
\begin{align} %
\left(\Psi_1\:,\Psi_2\right) =
\int\limits_{t=const}\sqrt{|g_{AdS_{d+1}}|} d^{d}x\
E^{\underline{0}}_{\alpha}\
\overline{\widetilde{\Psi}}_1\Gamma^\alpha\widetilde{\Psi}_2%
=\int\limits_{t=const}\sqrt{|g_{AdS_d}|} d^{d-1}x\
e^{\underline{0}}_{\alpha} \int\limits^{\frac \pi 2}_{-\frac \pi 2}
d\varphi\ \overline{\Psi}_1\Gamma^\alpha\Psi_2\:. \label{DiracNorm}%
\end{align}%
The norm conservation condition requires that solutions satisfy
the following boundary conditions at
$\varphi=\pm\frac\pi 2$
\be\label{Boundary_Dirac} %
\left. \int\sqrt{g_{AdS_d}}d^{d-1}x\
 \overline{\Psi}_1\Gamma^\bullet\Psi_2\right|_{\varphi=\pm\frac
\pi 2}=0\:.%
\ee%

The analysis for even and odd $d+1$ is different.
\subsubsection{Odd $d$}
For odd $d$ we use the following representation of gamma matrices
$\Gamma^\alpha$ in $d+1$ dimensions
\be %
\Gamma^\mu=\gamma^\mu\otimes\sigma^1\:,\quad %
\Gamma^\bullet=1\otimes i\sigma^2\:, %
\ee%
where $\{\gamma^\mu,\gamma^\nu\}=2\eta^{\mu\nu}$ and $\sigma^1$,
$\sigma^2$, $\sigma^3$ are Pauli matrices. Let us expand the field
$\Psi$ as follows
\be%
\Psi(x,\varphi)=\sum_{n}
\left(\psi^+_n(x)\otimes\chi_n^+(\varphi)+\psi^-_n(x)
\otimes\chi_n^-(\varphi)\right)\:,%
\ee%
where $\psi^\pm_n(x)$ are $o(d-1,2)$ Weyl spinors. The
substitution of this expansion into the action (\ref{Dirac_Action})
gives the sum of Dirac actions in $d$ dimensions
\be\label{DiracActionEven} %
S=S^++S^-\:, \quad%
S^\pm =\sum_{n} \int \Vol{d}
\overline{\psi}{}^\pm_n\left(i\gamma^\mu e^{\underline{\mu}}_\mu
D_{\underline{\mu}}\mp m_n \right)\psi_n^\pm\:,
\ee%
where $m_n$ are eigenvalues of the mass operator
\be%
\widehat{m} = \left( %
\begin{array}{ll}
i\pd_\varphi \quad 0\\
0 \quad -i\pd_\varphi %
\end{array}\right)\:
\ee%
$\widehat{m}\chi_n^\pm = \pm m_n\chi_n^\pm$. The eigenvectors
$\chi_n$ can be represented as
\be %
\chi^+_n(\varphi) =\frac{1}{\sqrt{\pi}} \left(%
\begin{array}{c}%
\cos\alpha\ e^{-im_n\varphi}\\%
\sin\alpha\ e^{im_n\varphi}%
\end{array}\right)\:,\quad%
\chi^-_n(\varphi) =\frac{1}{\sqrt{\pi}} \left(%
\begin{array}{c}%
-\sin\alpha\ e^{im_n\varphi}\\%
\cos\alpha\ e^{-im_n\varphi}%
\end{array}\right)\:,%
\ee%
where $\alpha\in\left[0,2\pi\right)$.

For any $\Psi^\pm_n(x,\varphi)=
\psi^\pm_n(x)\otimes\chi^\pm_n(\varphi)$ the scalar product
(\ref{DiracNorm}) is
\begin{align} %
\left(\Psi^\pm_n,\Psi^\pm_{n^\prime}\right) &=
\int\limits_{t=const}\sqrt{|g_{AdS_d}|}d^{d-1}x\ e^{\underline{0}}_\mu\ %
\overline{\psi}^\pm_n\gamma^\mu\psi^\pm_{n^\prime}\int\limits^{\frac
\pi 2}_{-\frac \pi 2}d\varphi\ (\chi^\pm_n)^\dagger\chi^\pm_{n^\prime} \\ %
&=\delta_{n\:n^\prime}\int\limits_{t=const}\sqrt{|g_{AdS_d}|}d^{d-1}x\
e^{\underline{0}}_\mu\
\overline{\psi}^\pm_n\gamma^\mu\psi^\pm_{n^\prime} %
\end{align} %
and
\be\nn %
\left(\Psi^\pm_n,\Psi^\mp_{n^\prime}\right) = 0\:,%
\ee%
provided that the mass spectrum is
\be\nn %
m_n = 2n + a\:,\quad n\in\mathbb{Z}\:,\ a\in\left[0,2\right). %
\ee%

The boundary conditions (\ref{Boundary_Dirac}) determine the values
of $a=\frac 1 2$ and $\alpha=\frac{\pi}{4}$. Therefore, we have the
spectrum
\be\label{MassDirac} %
m_n=2n+\frac 1 2\:,\quad n\in\mathbb{Z}\:.%
\ee%

These results are in agreement with the group-theoretical analysis
of Section \ref{sUnitRepr}. Namely, the energy of massless Dirac
field in $AdS_{d+1}$ is $E_0=\frac d 2$. The mass parameter in
$AdS_d$ is $m_n=E_n-\frac{d-1}{2}$. By Theorem \ref{Thr_fer}, the
degression of a massless spin 1/2 field results in two mass spectra
both defined by the formula $m_n=n+\frac 1 2$, $n=0,1,\ldots$ To
relate they with (\ref{DiracActionEven}) and (\ref{MassDirac}), we
rewrite the spectrum (\ref{MassDirac}) as $m_n=2n+\frac 1 2$ and
$-m_n=2n+1 + \frac 1 2$, $n=0,1,\ldots$
\subsubsection{Even $d$}

Let us now consider the case of even $d$. We expand Weyl $o(d,2)$ spinor $\Psi$ as
follows
\be\label{ExpansionEven} %
\Psi =\sum_n\left(\psi^+_n(x)f^+_n(\varphi)+\psi^-_n(x)f^-_n(\varphi)\right)\:,%
\ee%
where $f^\pm_n(\varphi)$ are complex functions,
$\psi^\pm_n\equiv\Pi^\pm\psi_n$ and $\Pi^\pm=\frac{1}{2}(1\pm
i\Gamma^\bullet)$ are projectors, \ie
\be %
i\Gamma^\bullet\psi^\pm_n = \pm\psi^\pm_n\:.%
\ee%
Substituting the expansion (\ref{ExpansionEven}) into the action
(\ref{Dirac_Action}), we demand that the resulting action be of the form
\be\label{DiracActionOdd} %
S = \sum_n \int\Vol{d}\overline{\psi}_n\left(i\Gamma^\mu
e^{\underline{\mu}}_\mu D_{\underline{\mu}}-m_n\right)\psi_n\:.
\ee%
This gives the equations
\be\label{DiracEvenSystem} %
\pd_\varphi f^+_n = - m_n f^-_n\:,\quad \pd_\varphi f^-_n =  m_n f^+_n%
\ee%
along with the normalizing conditions
\be\label{DiracEvenNorms}%
\int\limits^{\frac\pi 2}_{-\frac\pi 2} d\varphi\:
(f^\pm_n)^*f^\pm_{n^\prime} =\delta_{n\:n^\prime}\:.%
\ee%

From the conditions (\ref{DiracEvenNorms}) it follows that, for any
$\Psi_n(x,\varphi) = \psi^+_n(x)f^+_n(\varphi)
+\psi^-_n(x)f^-_n(\varphi)$, the scalar product (\ref{DiracNorm}) is
\be %
\left(\Psi_n\:,\Psi_{n^\prime}\right) = \delta_{n\:n^\prime}
\int\limits_{t=const}\sqrt{|g_{AdS_d}|} d^{d-1}x\
e^{\underline{0}}_{\alpha}\
\overline{\psi}_n\Gamma^\alpha\psi_{n^\prime}\:.%
\ee%

The boundary condition (\ref{Boundary_Dirac}) requires
\be\label{DiracBoundaryEven} %
\left.f^+_{n^\prime}(f^-_n)^*\right|_{\varphi=\pm\frac\pi 2}=0\:.%
\ee%

The complete orthonormal sets of functions
$f^\pm_n$ that satisfy (\ref{DiracEvenSystem}), (\ref{DiracEvenNorms}),
(\ref{DiracBoundaryEven}) are
\begin{align}%
f^+_n = \frac{1}{\sqrt{2\pi}}\Big(ie^{im_n\varphi}+e^{-im_n\varphi}\Big)\,,\qquad%
f^-_n = \frac{1}{\sqrt{2\pi}}\Big(ie^{-im_n\varphi}+e^{im_n\varphi}\Big)%
\end{align}%
with the mass spectrum
\be\label{d_even_mass}%
m_n=2n+\frac 1 2,\quad n\in\mathbb{Z}%
\ee%

Taking into account that a sign of $m_n$ does not matter we observe
that, in agreement with Section \ref{sUnitRepr},
this mass spectrum is equivalent to
\be%
m_n=n+\frac 1 2\:,\quad n=0,1,2,\ldots\,.%
\ee%

\subsection{Massless spin one}

The Maxwell action of a vector massless field $h^\alpha$ in
$AdS_{d+1}$ is
\be\nn%
S = -\frac 1 4\int\Vol{d+1}F_{\alpha\beta}F^{\alpha\beta},%
\ee%
where $F_{\alpha\beta}=\pd_\alpha h_\beta-\pd_\beta h_\alpha$. It is
invariant under the spin one gauge transformation
\be\nn%
\delta h_\alpha = \partial_\alpha\xi\: .%
\ee%

Using index conventions (\ref{IndexDecomposition}), the vector field
$h_\alpha$
in $d+1$ dimensions decomposes into a vector $h^\mu$ and a scalar
$h^\bullet$ in $d$ dimensions. In the coordinates
(\ref{MetricDecomposition}), the action takes the form
\begin{align}%
S=&\int\Vol{d}\int dz\cosh^d(z)\left\{ -\frac 1 4 F_{\mu\nu}F^{\mu\nu}\right. \\%
&-\frac 1 2 \cosh^2(z) h^\mu\left[\pd^2_z + (d+2)\tanh(z)\pd_z +
2d-\frac{2(d-1)}{\cosh^2(z)}\right]h_\mu \label{S1_mass} \\ %
&\left.+\frac{1}{2\cosh^2(z)}\pd_\mu h^\bullet
\pd^\mu h^\bullet + \frac{1}{\cosh^2(z)}\pd_\mu h^\bullet\pd_z \cosh^2(z)h^\mu\right\} \:. \label{S1_diag}%
\end{align}%
The gauge transformations of the $d$-dimensional fields are
\begin{align} %
\delta h^\mu &= \pd^\mu\xi\:,\label{S1GaugeIndexDecomp} \\ %
\delta h^\bullet &= -\pd_z\cosh^2(z)\xi\:. \nn %
\end{align}%

Let us expand the vector and scalar fields in $d$ dimensions as
\be %
\label{EXP1}
h^\mu = \sum^\infty_{n=0}a^\mu_n(x)P^1_n(z),\qquad%
h^\bullet = \sum^\infty_{n=0}\phi_n(x)R^1_n(z)\:,%
\ee%
where both the set of  functions $P^1_n(z)$ and the set $R^1_n(z)$
form  orthonormal  bases with respect to appropriate norms whose
precise  form will be specified below.
Eq.~(\ref{S1GaugeIndexDecomp}) suggests the expansion for the gauge
parameter
\be\nn%
\xi = \sum^\infty_{n=0}\xi_n(x)P^1_n(z)\:.%
\ee%

The basis functions can be chosen in such a way that the
substitution of the expansions (\ref{EXP1}) into the action gives
\be\label{VectorFieldLagrangian} %
S=\sum^\infty_{n=0}S_n,%
\ee%
\begin{align*}\nn%
S_n=\int\Vol{d}\left\{-\frac 1 4 f_{n\mu\nu}f_n^{\mu\nu} +
\frac{m^2_n}{2}a_n^\mu a_{n\mu} + m_n a^\mu_n\pd_\mu\phi_{n+1} +
\frac 1 2 \pd_\mu \phi_n \pd^\mu \phi_n\right\} \:,
\end{align*}%
where $f_{n\mu\nu}=\pd_\mu a_{n\nu}-\pd_\nu a_{n\mu}$ and the masses
are $m^2_n = (n+1)(n+d-2)$. The gauge transformations take the form
\be\label{S1_gauge} %
\delta a^\mu_n=\pd^\mu\xi_n\:, \quad %
\delta \phi_n=-m_{n-1}\xi_{n-1}\:,\quad n=0,1,2,\ldots%
\ee%

To obtain the action (\ref{VectorFieldLagrangian}) the following
properties have to be satisfied by $P^1_n(z)$
and $R^1_n(z)$. The diagonalization of the spin-one part
(\ref{S1_mass}) demands $P_n^1(z)$ to solve the equation
\be\label{S1_P}%
\pd^2_zP^1_n(z)+(d+2)\tanh(z)\pd_zP^1_n(z)+\left[\frac{m_n^2-2(d-1)}{\cosh^2(z)}+2d\right]P^1_n(z)=0.
\ee%
The diagonalization condition for the cross term (\ref{S1_diag})
gives the equation for $R^1_n(z)$
\be\label{S1_R}%
\pd^2_zR^1_n(z)+d\tanh(z)\pd_zR^1_n(z)+\left[\frac{M_n^2-d+2}{\cosh^2(z)}+2(d-2)\right]R^1_n(z)=0,
\ee%
where $M^2_n=n(n+d-3)$. The normalization conditions on $P^1_n(z)$
and $R^1_n(z)$ result from the integration  of the coefficients in front of the kinetic terms of
the action over the degression coordinate $z$
\be\label{NormS1} %
\int\limits^\infty_{-\infty}dz\cosh^d(z)P^1_n(z)P^1_m(z)=\delta_{nm}\:,%
\quad
\int\limits^\infty_{-\infty}dz\cosh^{d-2}(z)R^1_n(z)R^1_m(z)=\delta_{nm}\:.%
\ee%

The solutions of the equations (\ref{S1_P}), (\ref{S1_R}) and
(\ref{NormS1}) are
\be\nn %
P^1_n(z)=N^{\frac{d-2}{2}}_n\cosh^{-d}(z){}_2F_1\left(-n,n+d-1;\frac{d}{2};\frac{1+\tanh(z)}{2}\right),%
\ee%
\be\nn %
R^1_n(z)=N^{\frac{d-4}{2}}_n\cosh^{-d+2}(z){}_2F_1\left(-n,n+d-3;\frac{d-2}{2};\frac{1+\tanh(z)}{2}\right),%
\ee%
where $N^\kappa_n$ is defined in (\ref{NormFactor}). The $P^1_n(z)$
and $R^1_n(z)$ form orthonormal complete sets with respect to the
corresponding norms. Note that $P^1_n(z)$ and $R^1_n(z)$ are related
by
\be\nn%
\frac{1}{\cosh^{d-2}(z)}\pd_z\left(\cosh^{d-2}(z)R^1_{n+1}(z)\right)=-m_nP^1_n(z)\:.%
\ee%

The gauge transformation laws (\ref{S1_gauge}) for the scalar fields
$\phi_n$ with $n=1,2\ldots$ imply that all of them are Stueckelberg,
\ie can be gauged fixed to zero. As a result, the dimensional
degression of the massless spin one field gives the tower of massive
spin one fields with the energies
\be%
E_n=n+d-1\:,\quad n=0,1,\ldots
\ee%
and just one scalar field with energy $E=d-1$. This precisely
matches the pattern of Theorem \ref{ThSpin}. Let us note again that
all fields resulting from the dimensional degression of a massless
spin one are massive.

\subsection{Massless spin two}%

A massless field of spin two $h^{\alpha_1\alpha_2}$ in $AdS_{d+1}$
is described by the action \cite{Fronsdal:1978vb}
\be\begin{split}\label{LagrangianSpin2} %
S=&\int\Vol{d+1}\left\{ \frac 1 2 \nabla^\alpha
h_{\beta_1\beta_2}\nabla_\alpha h^{\beta_1\beta_2} - \nabla_\alpha
h^{\alpha\beta}\nabla^\gamma h_{\gamma\beta}\right. \\%
&\left. + \nabla_\alpha h^\prime\nabla_\beta h^{\alpha\beta} - \frac
1 2\nabla^\alpha h^\prime\nabla_\alpha h^\prime +
h_{\alpha_1\alpha_2}h^{\alpha_1\alpha_2} +
\frac{d-2}{2}h^\prime h^\prime\right\} \:, %
\end{split}\ee%
where
\be\nn %
h^\prime= g_{\alpha\beta}h^{\alpha\beta}.%
\ee%
The action (\ref{LagrangianSpin2}) is invariant under the gauge
transformation
\be\nn%
\delta h^{\alpha\beta}=\nabla^{(\alpha}\xi^{\beta)}\:,%
\ee%
where $\nabla^{\alpha}$ is the covariant derivative in $AdS_{d+1}$.

Using the index conventions (\ref{IndexDecomposition}) the field
$h^{\alpha_1\alpha_2}$ in $d+1$ dimensions decomposes into a spin
two field $h^{\mu_1\mu_2}$, a vector $h^{\mu\bullet}$, and a scalar
$h^{\bullet\bullet}$ in $d$ dimensions. In the coordinates
(\ref{MetricDecomposition}) the action (\ref{LagrangianSpin2}) takes
the form
\be\nn%
S=S_{22} + S_{22}^\prime+S_{22}^{\prime\prime}+S_{11} +S_{21}+S_{21}^\prime+S_0,%
\ee%
where
\begin{align}%
S_{22}&=\int\Vol{d}\int dz\cosh^d(z)\left\{ \frac 1 2 \cosh^2(z)D^\mu h_{\sigma_1\sigma_2}D_\mu%
h^{\sigma_1\sigma_2} - \cosh^2(z)D^\mu h_{\mu\sigma}D_\nu
h^{\nu\sigma} \right. \\ %
&\left.+\frac 1 2
\cosh^4(z)h_{\sigma_1\sigma_2}\left(\pd^2_z+(d+4)\tanh(z)\pd_z
+2d\tanh^2(z) + 4\right)h^{\sigma_1\sigma_2}\right\}\:, \label{S2_mass} %
\end{align}%
\be%
S_{22}^\prime =\int\Vol{d}\int dz\cosh^{d+2}(z) D_\nu
h^\prime D_\mu h^{\mu\nu}\:,\quad h^\prime\equiv g_{\mu\nu}h^{\mu\nu}\:, %
\ee%
\begin{align}%
S_{22}^{\prime\prime}&=\int\Vol{d}\int dz\cosh^{d+2}(z)\left\{-\frac
1 2 D^\mu
h^\prime D_\mu h^\prime +\frac{d-1}{2}h^\prime h^\prime\right. \\%
&\left.-\frac 1 2 \cosh^2(z)h^\prime\left(\pd^2_z+(d+4)\tanh(z)\pd_z
+2d\tanh^2(z) + 4\right)h^\prime\right\}\:,%
\end{align}%
\begin{align}%
S_{11}&=\int\Vol{d}\int dz\cosh^d(z)\left\{-D^\mu h_\sigma^\bullet
D_\mu h^{\sigma\bullet}+D_\mu h^{\mu\bullet}D_\nu
h^{\nu\bullet} +(d-1)h_\sigma^\bullet h^{\sigma\bullet}\right\}\:, %
\label{S2_S1xS2} %
\end{align}%
\be%
S_{21}=-2\int\Vol{d}\int dz\cosh^{d+2}(z) D^\mu
h_{\mu\nu}\frac{1}{\cosh^d(z)}\pd_z\cosh^d(z)h^{\nu\bullet}\:, %
\ee%
\be%
S_{21}^\prime=2\int\Vol{d}\int dz\cosh^d(z)D_\mu
h^{\mu\bullet}\pd_z\cosh^2(z)h^\prime\:, %
\ee%
\be\begin{split}\label{S2_S0} %
S_0&=\int\Vol{d}\int dz\cosh^d(z)\left\{h_{\bullet\bullet}D_\mu
D_\nu h^{\mu\nu} + D^\mu
h^\prime D_\mu h_{\bullet\bullet}+\frac{d(d-1)}{2}\tanh^2(z)h_{\bullet\bullet}^2 \right.\\
&\left.+(d-1)h_{\bullet\bullet}\tanh^2(z)\pd_z\cosh^2(z)\coth(z)h^\prime
+2(d-1)\tanh(z)h_{\bullet\bullet}D_\mu
h^{\mu\bullet}\vphantom{\frac{d(d-1)}{2}}\right\}\:. %
\end{split}\ee%

The gauge transformations take the form
\begin{align}%
\delta h^{\mu\nu}&=\frac 1 2 \left(D^\mu \xi^\nu + D^\nu \xi^\mu\right) + \tanh(z)\xi g^{\mu\nu}\:,\label{g_s2} \\ %
\delta h^{\mu\bullet}&= -\frac 1 2\pd_z\cosh^2(z)\xi^\mu + \frac 1 2 D^\mu\xi\:,\label{g_s1} \\%
\delta h_{\bullet\bullet}&= - \pd_z\cosh^2(z)\xi\:.\label{g_s0} %
\end{align}%

A new feature of the spin two case   compared to spins zero and one
is that there are two types of scalar field modes. One comes from
the scalar component $h_{\bullet\bullet}$ while another one comes
from the trace part of $h^{\mu\nu}$ which had no analogue for $s=0$
or $1$. This results in the ambiguity in field redefinitions that
mix $h_{\bullet\bullet}$ with $h^\mu{}_{\mu}$. To obtain the action in
the form of infinite sum of actions that describe finite subsystems
of fields, the field variables should be chosen as follows
\begin{align}%
\widetilde{h}^{\mu\nu} &= h^{\mu\nu} - \frac{1}{d-2}g^{\mu\nu}\cosh^{-2}(z)h_{\bullet\bullet},\label{ten_redef} \\ %
\widetilde{h}_{\bullet\bullet} &= h_{\bullet\bullet}\:.\nn %
\end{align}%
In the sequel, we shall discard tilde over the new fields
$\widetilde{h}^{\mu\nu},\ \widetilde{h}_{\bullet\bullet}$ denoting
them as $h^{\mu\nu},\ h_{\bullet\bullet}$.

The redefinition (\ref{ten_redef}) only affects $S_0$ (\ref{S2_S0})
as the only one containing the field $h_{\bullet\bullet}$. It gets
the form
\begin{align*}%
\tilde{S}_0&=\int\Vol{d}\int dz\cosh^d(z)\left\{\frac 1
2\frac{d-1}{d-2}\cosh^{-2}(z)D_\mu
h_{\bullet\bullet}D^\mu h_{\bullet\bullet} +\right. \\%
&+\frac{d}{2}\frac{d-1}{(d-2)^2}h_{\bullet\bullet}\left(-\pd_z^2+(d-4)\tanh(z)\pd_z+(d-1)(d-2)\tanh^2(z)-d+2\right)h_{\bullet\bullet}-\\%
&-\frac{d-1}{d-2}\cosh^2(z)h^\prime\left(\pd_z^2+2(d-1)\tanh(z)\pd_z+(d-1)(d-2)\tanh^2(z)+d-2\right)h_{\bullet\bullet}+\\ %
&\left.+2\frac{d-1}{d-2}D_\mu
h^{\mu\bullet}\frac{1}{\cosh^{d-2}(z)}\pd_z\cosh^{d-2}(z)h_{\bullet\bullet}\right\}\:. %
\end{align*}%
In terms of new fields, the gauge transformations
(\ref{g_s2}),(\ref{g_s1}),(\ref{g_s0}) read as
\begin{align}%
\delta h^{\mu\nu}&=\frac 1 2 \left(D^\mu \xi^\nu + D^\nu \xi^\mu\right) +%
\frac{1}{d-2}g^{\mu\nu}\cosh^{-d}(z)\pd_z\cosh^d(z)\xi\label{Gauge_S2}\:, \\ %
\delta h^{\mu\bullet}&= -\frac 1 2\pd_z\cosh^2(z)\xi^\mu + \frac 1 2 D^\mu\xi\label{Gauge_S1}\:, \\%
\delta h_{\bullet\bullet}&= - \pd_z\cosh^2(z)\xi\:. \label{Gauge_S0}%
\end{align}%

Let us expand the tensor, vector, and scalar fields as follows
\be\nn%
h^{\mu_1\mu_2}=\sum^\infty_{n=0}\phi^{\mu_1\mu_2}_n(x)P^2_n(z),\quad%
h^{\mu\bullet}=\sum^\infty_{n=0}\phi^{\mu}_n(x)R^2_n(z),\quad%
h_{\bullet\bullet}=\sum^\infty_{n=0}\phi_n(x)Q^2_n(z),%
\ee%
where the sets of functions $\{P^2_n(z)\}$, $\{R^2_n(z)\}$, and $\{Q^2_n(z)\}$
form orthogonal bases with respect to appropriate norms
whose specific form will be given later on.
The gauge transformations (\ref{Gauge_S2}) and (\ref{Gauge_S1})
suggest analogous expansion for the gauge parameters
\be\nn%
\xi^\mu = \sum^\infty_{n=0}\xi^\mu_nP^2_n(z),\quad
\xi=\sum^\infty_{n=0}\xi_nR^2_n(z)\:.%
\ee%

The idea is choose the functions $P^2_n(z)$, $R^2_n(z)$, and
$Q^2_n(z)$ such that the resulting action acquire the form of the sum
of actions of spin two, spin one, and spin zero fields plus some
lower-derivative cross-terms that mix different fields. The
resulting action takes the form
\be\label{LagrangianS2} %
S=S_{2}+S_{11}+S_{21}+S_{21}^\prime+\tilde{S}_0\:,%
\ee%
where
\begin{align*}%
S_2&=\sum^\infty_{n=0}\int\Vol{d}\left\{\frac 1 2 D^\mu\phi_{n\sigma_1\sigma_2}%
D_\mu \phi^{\sigma_1\sigma_2}_n - D^\mu \phi_{n\mu\sigma}D_\nu
\phi_n^{\nu\sigma}+D_\nu \phi_n^\prime D_\mu \phi_n^{\mu\nu}\right.\\%
&-\left.\frac 1 2 D^\mu \phi_n^\prime D_\mu \phi_n^\prime +
\frac{m^2_n+d-1}{2} \phi_n^\prime \phi_n^\prime-%
\frac{m_n^2}{2}
\phi_{n\sigma_1\sigma_2}\phi_n^{\sigma_1\sigma_2}\right\}\:,\quad
\phi_n^\prime\equiv g_{\mu\nu}\phi_n^{\mu\nu}\:, %
\end{align*}%
\begin{align*}%
S_{11}&=\sum^\infty_{n=0}\int\Vol{d}\left\{%
-D^\mu \phi_{n\nu}D_\mu \phi^{\nu}_n+D_\mu \phi_n^{\mu}D_\nu
\phi^{\nu}_n + (d-1)\phi_{n\mu}\phi_n^\mu\vphantom{\frac 1 2}\right\}\:,\\ %
S_{21}&=\sum^\infty_{n=0}2\sqrt{(n+1)(n+d)}\int\Vol{d}D^\mu\phi_{n\mu\nu}\phi^\nu_{n+1}\:,\\ %
S_{21}^\prime&=\sum^\infty_{n=0}2\sqrt{(n+1)(n+d)}\int\Vol{d}\phi^\prime_n D_\mu\phi^\mu_{n+1}\:,\\ %
\tilde{S}_0 &=
\frac{d-1}{d-2}\sum^\infty_{n=0}\int\Vol{d}\left\{\frac
1 2 D_\mu\phi_nD^\mu\phi^n+\frac{d}{2}\frac{(n-1)(n+d-2)}{d-2}\phi^2_n \right.\\ %
&\left.-\sqrt{(n+1)(n+2)(n+d)(n+d-1)}\phi_n^\prime\phi_{n+2} -
2\sqrt{(n+1)(n+d-2)}\phi_{n+1}D_\mu\phi^\mu_n\vphantom{\frac 1
2}\right\}\:.
\end{align*}%
The gauge transformations are
\begin{align}%
\delta \phi_n^{\mu\nu}&=\frac 1 2 \left(D^\mu \xi_n^\nu + D^\nu \xi_n^\mu\right) %
+ \frac{\sqrt{(n+1)(n+d)}}{d-2}\xi_{n+1} g^{\mu\nu}\:,\label{S2GaugeS2} \\ %
\delta \phi^\mu_n&= -\frac 1 2\sqrt{n(n+d-1)}\xi_{n-1}^\mu + \frac 1 2 D^\mu\xi_n\:,\label{S2GaugeS1} \\%
\delta \phi_n&= -\sqrt{n(n+d-3)}\xi_{n-1}\:.\label{S2GaugeS0}%
\end{align}%

To bring the action (\ref{LagrangianS2}) to the desired form the
following conditions on the basis functions have to be imposed. The
diagonalization condition  of the spin-two part (\ref{S2_mass})
gives the following  equation on $P^2_n(z)$
\be\label{S2_P} %
\pd^2_zP^2_n(z)+(d+4)\tanh(z)\pd_zP^2_n(z)+\left[\frac{m^2_n-2d}{\cosh^2(z)}+2d+4\right]P^2_n(z)=0\:,
\ee%
where
\be%
m^2_n=n^2+(d+1)n+d-2\:%
\ee%
is the mass square parameter in the action. The diagonalization
condition for the vector field part (\ref{S2_S1xS2}) of the action
and gauge symmetry (\ref{Gauge_S0}) gives the equations on
$R^2_n(z)$ and $Q^2_n(z)$
\be\label{S2_R} %
\pd^2_zR^2_n(z)+(d+2)\tanh(z)\pd_zR^2_n(z)+\left[\frac{M^2_{R\:n}-d}{\cosh^2(z)}+2d\right]R^2_n(z)=0\:,
\ee%
\be\label{S2_Q} %
\pd^2_zQ^2_n(z)+d\tanh(z)\pd_zQ^2_n(z)+\left[\frac{M^2_{Q\:n}-d+2}{\cosh^2(z)}+2(d-2)
\right]Q^2_n(z)=0\:, %
\ee%
where
\begin{align}%
M^2_{R\:n}&=n(n+d-1)\:,\\ %
M^2_{Q\:n}&=n(n+d-3)\:,\quad n=0,1,\ldots %
\end{align}%

The normalization conditions on $P^2_n(z),R^2_n(z)$, and $Q^2_n(z)$
are obtained from the integral over the degression coordinate $z$ in
the corresponding kinetic terms of the action
\be\label{NormSpin2_1} %
\int\limits^\infty_{-\infty}dz\cosh^{d+2}(z)P^2_n(z)P^2_m(z)=\delta_{nm}\:,%
\quad %
\int\limits^\infty_{-\infty}dz\cosh^{d}(z)R^2_n(z)R^2_m(z)=\delta_{nm}\:,%
\ee%
\be\label{NormSpin2_2} %
\int\limits^\infty_{-\infty}dz\cosh^{d-2}(z)Q^2_n(z)Q^2_m(z)=\delta_{nm}\:.%
\ee%
The solutions of equations (\ref{S2_P}),(\ref{S2_R}),(\ref{S2_Q})
that satisfy the normalizablility conditions (\ref{NormSpin2_1}),
(\ref{NormSpin2_2}) are
\begin{align} %
P^2_n(z)&=N^{\frac d
2}_n\cosh^{-d-2}(z){}_2F_1\left(-n,n+d+1;\frac{d+2}{2};\frac{1+\tanh(z)}{2}\right)\:,\\ %
R^2_n(z)&=N^{\frac{d-2}{2}}_n\cosh^{-d}(z){}_2F_1\left(-n,n+d-1;\frac{d}{2};\frac{1+\tanh(z)}{2}\right)\:,\\ %
Q^2_n(z)&=N^{\frac{d-4}{2}}_n\cosh^{-d+2}(z){}_2F_1\left(-n,n+d-3;\frac{d-2}{2};\frac{1+\tanh(z)}{2}\right)\:,%
\end{align} %
where $N_n^\kappa$ is defined by (\ref{NormFactor}).
The following relationships between $P^2_n$, $R^2_n$, and $Q^2_n$ hold
\begin{align*}%
\cosh^{-d}(z)\pd_z\left(\cosh^{d}(z)R^2_n(z)\right)&=-M_{R\:n}P^2_{n-1}(z)\:, \\%
\pd_z\left(\cosh^2(z)P^2_n(z)\right)&=M_{R\:n+1}R^2_{n+1}(z)\:, \\%
\pd_z\left(\cosh^2(z)R^2_n(z)\right)&=M_{Q\:n+1}Q^2_{n+1}(z)\:, \\ %
\cosh^{-d+2}(z)\pd_z\left(\cosh^{d-2}(z)Q^2_n(z)\right)&= -M_{Q\:n}R^2_{n-1}(z)\:. %
\end{align*}%

We observe that all gauge symmetries (\ref{S2GaugeS2}),
(\ref{S2GaugeS1}) and (\ref{S2GaugeS0}) are Stueckelberg. Upon gauge
fixing the Stueckelberg fields to zero, we find that the spectrum of
fields resulting from the dimensional degression contains  a tower
spin two fields with masses and energies
\be\nn%
m^2_n=n^2+(d+1)n+d-2\:,\quad E_n=n+d\:,\quad n=0,1,\ldots%
\ee%
a single vector field $\phi^\mu_0$ and a single scalar field
$\phi_0$ with masses $d-1$, and $d$, respectively, both having the
 energy $E_0=d$. This precisely matches Theorem
\ref{ThSpin}.

\section{Scalar field degression over several dimensions}
\label{sRedMoreDim}

Let $AdS_{d+\ad}$ with $\ad\geq 2$ be realized as a hyperboloid
\be\nn%
y_{-1}^2+y_0^2-y_1^2-\ldots-y_{d-1}^2
-y_d^2-\ldots-y^2_{d+\ad-1}=1 %
\ee%
in $\mathbb{R}^{d+\ad+1}$. We choose the following coordinates
\be\nn%
y_{-1}=\cosh(z)\tilde{y}_{-1},\
y_{0}=\cosh(z)\tilde{y}_{0},\:\ldots\:,\
y_{d-1}=\cosh(z)\tilde{y}_{d-1},%
\ee%
\be\nn%
y_d=\sinh(z) \tilde{y}_d,\:\ldots\:,\ y_{d+\ad-1}=\sinh(z) \tilde{y}_{d+\ad-1},%
\ee%
where $z\in [0,\infty)$,
$\tilde{y}_{-1},\ldots,\tilde{y}_{d-1}$ describe a hyperboloid
\be\nn%
\tilde{y}_{-1}^2+\tilde{y}_0^2-\tilde{y}_1^2-\ldots-\tilde{y}_{d-1}^2=1
\ee%
in a $\mathbb{R}^{d+1}$, while $\tilde{y}_d,\ldots,\tilde{y}_{d+\ad-1}$
describe $S^{\ad-1}$
\be\nn%
\tilde{y}_d^2+\ldots+\tilde{y}^2_{d+\ad-1}=1.%
\ee%
 The line element of $AdS_{d+\ad}$ is
\be\nn%
ds^2_{AdS_{d+\ad}}=\cosh^2(z)ds^2_{AdS_d}-dz^2-\sinh^2(z)ds^2_{S^{\ad-1}}.%
\ee%

In this coordinates, the Laplace-Beltrami operator is
\begin{align*}%
\Box_{AdS_{d+\ad}} &= \frac{1}{\cosh^2(z)}\Box_{AdS_d}\\
&-\frac{1}{\cosh^d(z)\sinh^{\ad-1}(z)}\pd_z\cosh^d(z)\sinh^{\ad-1}(z)\pd_z
- \frac{1}{\sinh^2(z)}\Box_{S^{\ad-1}},
\end{align*} %
where $\Box_{AdS_d}$ and $\Box_{S^{\ad-1}}$ are the Laplace-Beltrami
operators on $AdS_d$ and $S^{\ad-1}$, respectively.

The free scalar field equation in $AdS_{d+\ad}$ is
\be\label{RedMoreDimAdSEq}%
(\Box_{AdS_{d+\ad}}+M^2)\Phi(x,z,u)=0,%
\ee%
where $x$ are local coordinates of $AdS_d$ and $u$ are those of
$S^{\ad-1}$. Let us look for its solutions in the form
\be\label{RedMoreDimFourier}%
\Phi(x,z,u) = \sum \limits^\infty_{N,L=0}\sum \limits_K
\phi_{NLK}(x)P_{NL}(z)Y_{LK}(u),
\ee%
where $Y_{LK}(u)$ are spherical functions on $S^{\ad-1}$,
$K=(k_1,\ldots,\pm k_{\ad-2})$ so that $L\geq k_1\geq\ldots\geq
k_{\ad-2}\geq 0$ and
\be\nn%
\Box_{S^{\ad-1}}Y_{LK}(u)=-L(L+\ad-2)Y_{LK}(u).%
\ee%
The functions $P_{NL}(z)$ form an orthogonal complete set
with respect to the norm
\be\label{RedMoreDimNorm}%
\int\limits^\infty_0dz\cosh^{d-2}(z)\sinh^{\ad-1}(z)P_{NL}(z)P_{N^\prime L}(z)=\delta_{NN^\prime} %
\ee%
inherited from (\ref{FieldTheorNorm}).
Requiring  $P_{NL}(z)$ to satisfy the equation
\begin{align}%
\pd_z^2 P_{NL}(z)+&\left(d\tanh(z)+(\ad-1)\coth(z)\right)\pd_zP_{NL}(z)\\%
&+\left(\frac{m_{NL}^2}{\cosh^2(z)}-\frac{l(l+\ad-2)}{\sinh^2(z)}-M^2\right)
P_{NL}(z)=0\nn
\end{align}%
and plugging the expansion
(\ref{RedMoreDimFourier}) into Eq.~(\ref{RedMoreDimAdSEq}), we
obtain the tower of massive scalar fields
\be\label{RedMoreDimScalarEq}%
(\Box_{AdS_d}+m_{NL}^2)\phi_{NLK}(x)=0\,.%
\ee%

The requirement of finiteness of the norm (\ref{RedMoreDimNorm})
constrains solutions $P_{NL}(z)$ to the ``irregular'' ones
\be\nn%
P^{irr}_{NL}(z)=N_{NL}^{-\kappa}\cosh^\Lambda(z)\sinh^{2N+L}(z) %
{}_2F_1\left(-N,-N-L-\frac \ad 2+1;1-\kappa;-\frac{1}{\sinh^2z}\right), %
\ee%
\be\nn %
\Lambda=-2N-L-\frac{d+\ad-1}{2}+\kappa\:,\quad%
M^2\in\left[-\frac{(d+\ad-1)^2}{4},-\frac{(d+\ad-1)^2}{4}+1\right)%
\ee%
and ``regular'' solutions
\be\nn%
P^{reg}_{NL}(z)=N_{NL}^\kappa\cosh^{\Lambda^\prime}(z)\sinh^{2N+L}(z)
{}_2F_1\left(-N,-N-L-\frac \ad
2+1;1+\kappa;-\frac{1}{\sinh^2z}\right),
\ee%
\be\nn %
\Lambda^\prime=-2N-L-\frac{d+\ad-1}{2}-\kappa\:,\quad %
M^2\in\left[-\frac{(d+\ad-1)^2}{4},+\infty\right)\:,%
\ee%
where $\kappa\equiv\sqrt{M^2+\frac{(d+\ad-1)^2}{4}}$ and
\be\label{NormaFactorMore}%
N_{NL}^\kappa=\Gamma(1+\kappa)\sqrt{\frac{N!\:\Gamma(N+L+\frac \ad
2)} {2\left(2N+L+\frac \ad
2+\kappa\right)\Gamma(\kappa+N+1)\Gamma(N+L+\frac \ad 2+\kappa)}} %
\ee%
is fixed by (\ref{RedMoreDimNorm}).

The lowest energy and mass spectra
are
\begin{align}%
E_{NL}&=2N+L+\frac{d+\ad-1-2\kappa}{2}\label{RedMoreDimEnergSing}\:,\\ %
m^2_{NL}&=\left(2N+L-\kappa+\frac \ad
2\right)^2-\frac{(d-1)^2}{4}\label{RedMoreDimMassaSing}%
\end{align}
for the ``irregular'' solutions and
\begin{align}%
E_{NL}&=2N+L+\frac{d+\ad-1+2\kappa}{2}\:,\label{RedMoreDimEnergReg}\\ %
m^2_{NL}&=\left(2N+L+\kappa+\frac \ad
2\right)^2-\frac{(d-1)^2}{4}\label{RedMoreDimMassaReg}\:%
\end{align}%
for the ``regular'' ones. Recall that the relation of the mass and the
energy is given by (\ref{MassEnergy}).

These results can also be lifted to the action level.
Substituting the expansion (\ref{RedMoreDimFourier}) into the
action
\be\nn%
S^{AdS_{d+\ad}}=\frac 1 2\int\Vol{d+\ad}\Phi(\Box_{AdS_{d+\ad}}+M^2)\Phi\:,%
\ee%
 we find that
\be\nn%
S^{AdS_{d+\ad}}=\sum\limits^\infty_{N=0,L=0}\sum_{K}S_{NLK}^{AdS_d},%
\ee%
where
\be\nn %
S_{NLK}^{AdS_d}=\frac 1 2\int\Vol{d}\phi_{NLK}(\Box_{AdS_d}+m^2_{NL})\phi_{NLK}.%
\ee%

These results are similar to those of the dimensional degression
over one coordinate. Recall that the appearance of ``irregular'' and
``regular'' solutions results from the ambiguity in the relation
between mass and energy. Fixing a particular vacuum energy $E_0$
value, we see that the dimensional degression over several
coordinates gives rise to a set of scalar fields of energies
$E_0+k$, $k=0,1,\ldots$ Each energy level has the multiplicity
(\ref{UnitMore}) in  agreement with the group-theoretical
analysis of Section \ref{sUnitRepr}

As was pointed out by Metsaev \cite{Metsaev:2000qb}, the dimensional
degression over two coordinates produces a massless scalar field
from a massless one. Namely, there are two energies
$E_0=\frac{d+\ad}{2}$ and $E_0=\frac{d+\ad-2}{2}$ corresponding to
the massless scalar field in $AdS_{d+\ad}$. If $\ad=2$ the second
energy becomes $E_0= \frac d 2$, \ie the energy of the massless
scalar field in $AdS_d$. Obviously, for spin $s\geq \frac 1 2$ this
phenomenon does not take place because there is only one energy
corresponding to a massless field. Thus, except for the scalar field
the dimensional degression of a massless field never gives a
massless field at least if the original field corresponds to some
symmetric representations of the $AdS$ algebra.

Let us compare our results with those of Metsaev
\cite{Metsaev:2000qb}. In the case of dimensional degression of a
massless scalar field over several coordinates  the mass spectrum
(\ref{RedMoreDimMassaSing}) for ``irregular'' solutions at $N=0$
corresponds to the spectrum given by the formula (47) in
\cite{Metsaev:2000qb}. The mass spectrum (\ref{RedMoreDimMassaReg})
for ``regular'' solutions corresponds to the spectrum given by
formula (54) of \cite{Metsaev:2000qb} but with some disagreement in
the restrictions (55) of \cite{Metsaev:2000qb} on the possible
values of the parameters $L$(denoted in \cite{Metsaev:2000qb} as
$l$) and $\kappa$. The difference is that the dimensional degression
of the scalar field with the mass in the interval
$-(d+\ad-1)^2/4\leq M^2< -(d+\ad-1)^2/4+1$ gives two mass spectra
(\ref{RedMoreDimMassaSing}) and (\ref{RedMoreDimMassaReg})
compatible with unitarity. The spectrum associated with
``irregular'' solutions is most likely ruled out in
\cite{Metsaev:2000qb} by too strong boundary conditions.

\section{Conclusion}
In this paper we extend the previously known results on the
dimensional degression of a massless scalar in anti-de Sitter space
to the cases of  massive scalar (including shadow sector) and
massless fields of spins one-half, one, and two. Our
field-theoretical analysis matches the group-theoretical analysis of
the branching of unitary $o(d+\ad,2)$--modules into $o(d,2)\oplus
o(\ad)$--modules also performed in this paper. It is shown that
dimensional degression of a massless field of spin $s\geq \frac 1 2$
in $d+1$ dimension produces an infinite set of massive fields in
lower dimension. The spectrum is discrete, being quantized in values
of the inverse $AdS$ radius. For this reason, in the $AdS$ case, it
is not necessary  to compactify space-time to obtain discrete
spectrum in lower dimension. (In fact it is not even clear what
could be an analog of the torus compactification in the $AdS$ case.)

Our results may have several applications. First of all this is the
first step towards the analysis of the dimensional
compactification-like effects in HS theories, that have $AdS$ rather
than Minkowski space-time as their most symmetric vacuum. This
analysis is interesting, in particular, for better understanding of
possible mechanisms of spontaneous breakdown of HS gauge theories.
Indeed, we have shown that the dimensional degression gives rise to
the gauge invariant (\ie Stueckelberg) formulation of massive
fields. As such it is just analogous to the standard torus
compactification mechanism used to derive massive HS theories from
the massless ones in one higher dimension \cite{SS,RSS}.

In other words, the degression mechanism in $AdS$ uncovers the
structure of necessary Higgs fields in HS gauge theory. Of course
our results for spin one and two are not new in this respect. The
gauge invariant formulation of symmetric fields has been now well
understood in the formalism of symmetric tensors
\cite{Zinoviev:2001dt,Buchbinder:2006ge,
Metsaev:2006zy,Buchbinder:2006nu}. However, to make it
appropriate for the analysis of nonlinear HS theories we should
better understand how the higgsing works in terms of the HS
frame-like formalism of \cite{Lopatin:1987hz,Vasiliev:1987tk,5d,SV}
\footnote{See also  recent papers \cite{SORV} and \cite{mf} that extend the frame-like
formalism to the cases of reducible sets of massless fields and massive
fields, respectively.} which is at the moment the only working one at the full interacting
level (see \cite{solv,SSS,sor} for more detail and references on nonlinear
HS theories). Moreover, an extension of our analysis to the case of
mixed symmetry massless fields in $AdS_d$ \cite{ASV}
will allow us to achieve a gauge invariant formulation of massive
fields of general symmetry type in $AdS_d$.

\section*{Acknowledgments}
The authors acknowledge with gratitude the collaboration of
O.~Shaynkman at the early stage of this work. We are grateful to
R.~R.~Metsaev for useful discussions and to V.~I.~Ritus for useful
remarks. This research was supported in part by INTAS Grant No
05-1000008-7928, RFBR Grant No 08-02-00963, LSS No 1615.2008.2
The work of M.V. was partially supported by the Alexander von
Humboldt Foundation Grant PHYS0167.
 The work of A.A. was partially supported by the
grant of Dynasty Foundation.

\section{Appendix}%
\setcounter{Thr}{0}

\begin{Thr}%
For an integer spin $s$, the $o(d,2)$-module $\Verma{E_0}{s}{d}$
with $E_0\geq E_0({\bf s})$ has the following branching into
$o(d-1,2)$--modules
\be \label{decApp}
\Verma{E_0}{s}{d}\simeq \sum^s_{q=0}\sum^\infty_{k=0}\oplus \Verma{E_0+k}{q}{d-1} .%
\ee%
\end{Thr}%

\begin{prf}%

An $o(d,2)$-module $\Verma{E_0}{s}{d}$ is spanned by vectors
(\ref{states})
\be\nn%
T^{+a_1}\ldots T^{+a_M} \left|E_0,{\bf s}\right>%
\ee%
with various $M$ and $a=1,\dots,d$. The irreducible vacuum
$o(d)\oplus o(2)$--module $\left|E_0,s\right>$, where $E_0$
and ${\bf s}=(s,0,\ldots,0)$ are  the
weights for $o(2)$  and $o(d)$, respectively,
decomposes as an $o(d-1)\oplus o(2)$-module into
\be\nn%
\sum^s_{q=0}\oplus\left|E_0,q\right>.%
\ee%
Using the index decomposition $n = 1,\ldots,d-1\:,\; \bullet = d$
and denoting $t^+\equiv T^{+\bullet}$, the module
$\Verma{E_0}{s}{d}$ is spanned by the vectors
\be%
T^{+n_1}\ldots T^{+n_N}(t^+)^k\left|E_0,q\right>\,,%
\ee%
where $N,k=0,1,\ldots \infty,\ q=0,\ldots,s$.

Let $W_{kq}$ be a vector space spanned by %
\be\nn%
T^{+n_1}\ldots T^{+n_N}(t^+)^k\left|E_0,q\right>\:,\quad N=0,1,\ldots%
\ee%
{}From the definitions of $W_{kq}$ it follows that%
\be\label{UnitWMinus_bos} %
T^{-n}W_{kq}\subset W_{kq}\oplus W_{k-1\:q-1}\oplus
W_{k-1\:q+1}\oplus
W_{k-2\:q}\:,%
\ee%
\be\label{UnitWPlus_bos} %
T^{+n}W_{kq}\subset W_{kq}\:,\quad T^{nm}W_{kq}\subset W_{kq}
\:,\quad EW_{kq}\subset W_{kq}\,. %
\ee%

Let $V_{ij}$ be vector spaces defined recurrently as
\be\nn%
V_{ij}=\sum^j_{q=0}\oplus W_{iq}\oplus V_{i-1\:s} \:, %
\ee%
where $i=0,1,\ldots, \infty$, $j=0,1,\ldots,s$, $V_{-1j}=\{0\}$.
According to this definition,
\be%
V_{ij}\subset V_{ij+1}\,,\qquad V_{ij}\subset V_{i+1\:j}\,.%
\ee%

From (\ref{UnitWMinus_bos}) and (\ref{UnitWPlus_bos}) it follows
that $V_{ij}$ are $o(d-1,2)$--modules. As a result we obtain the
filtration of the $o(d-1,2)$--module $\Verma{E_0}{s}{d}$ by the
$o(d-1,2)$--modules $V_{ij}$
\be\nn%
\{0\}\subset V_{00}\subset V_{01}\subset\ldots\subset V_{0s}\subset
V_{10}\subset\ldots = \Verma{E_0}{s}{d} \,.%
\ee%

Obviously, the composition factors form the following
$o(d-1,2)$--modules%
\be\nn%
V_{ij}/V_{ij-1}\simeq \Verma{E_0+i}{j}{d-1}\:,\quad j\neq 0\:, %
\ee%
\be\nn%
V_{i0}/V_{i-1s}\simeq \Verma{E_0+i}{0}{d-1}\,, %
\ee%
which are irreducible because the inequalities
(\ref{UnitCondLowest}) and (\ref{UnitCondS}) are satisfied in $d$
dimensions as a consequence of those in $d+1$ dimensions.
Here we use the unitarity of $\Verma{E_0}{s}{d}$ with
$E_0\geq E_0({\bf s})$ as an $o(d-1,2)$ module.
Clearly, once the inequality $E_0>E_0({\bf s})$ holds,
$\Verma{E_0}{s}{d}$ is a unitary $o(d,2)$--module and hence also
a unitary $o(d-1,2)$--module.
Taking into account that a reducible unitary module is fully
reducible ({\it i.e.} it decomposes into direct sum of irreducible
submodules), we conclude that (\ref{decApp}) is true.

The case where  the energy $E_0$ belongs to the boundary of the
unitary region $E_0=E_0({\bf s})$ is considered analogously,
eventhough, in this case, $\Verma{E_0({\bf s})}{s}{d}$ is not a
unitary $o(d,2)$--module because it contains  null  states that have
zero $o(d,2)$ invariant norm. The key observation is that the
energies of the lowest energy states of the $o(d-1,2)$--modules
contained in the boundary $o(d,2)$--modules remain inside the
unitarity region for $o(d-1,2)$. As a result, all
$o(d-1,2)$--modules remain irreducible on the boundary of the
unitarity region for $o(d,2)$. (Note, that this implies that one can
introduce an $o(d-1,2)$ invariant norm such that $\Verma{E_0({\bf
s})}{s}{d}$ be a unitary $o(d-1,2)$--module.)

\end{prf}

\begin{Thr}%
For a half-integer spin $s$ and $E_0\geq E_0({\bf s})$, the
$o(d,2)$-module $\Verma{E_0}{s^\pm}{d}$ for even $d$ and
$\Verma{E_0}{s}{d}$ for odd $d$  have, respectively, the following
branchings into $o(d-1,2)$--modules
\be \label{dec_even_app}
\Verma{E_0}{s^\pm}{d}\simeq \sum^{s}_{q=\frac 1 2}\sum^\infty_{k=0}\oplus \Verma{E_0+k}{q}{d-1},%
\ee%
\be \label{dec_odd_app}%
\Verma{E_0}{s}{d}\simeq \sum^{s^+}_{q^+=\frac 1
2}\sum^\infty_{k=0}\oplus \Verma{E_0+k}{q^+}{d-1}
\oplus \sum^{s^-}_{q^-=\frac 1 2}\sum^\infty_{k=0}\oplus \Verma{E_0+k}{q^-}{d-1}.%
\ee%
\end{Thr}%

\begin{prf}%

A unitary $o(d,2)$-module $\Verma{E_0}{s^\pm}{d}$ if $d$ is even and
$\Verma{E_0}{s}{d}$ if $d$ is odd is spanned by vectors
(\ref{states})
\be\nn%
T^{+a_1}\ldots T^{+a_M} \left|E_0,{\bf s}^\pm\right>\ \text{or} \quad %
T^{+a_1}\ldots T^{+a_M} \left|E_0,{\bf s}\right> %
\ee%
with various $M$. The irreducible vacuum $o(d)\oplus o(2)$--modules
$\left|E_0,s^\pm\right>$ and $\left|E_0,s\right>$ decompose as an
$o(d-1)\oplus o(2)$-modules into
\be\nn%
\sum^{s}_{q=\frac 1 2}\oplus\left|E_0,q\right>,%
\ee%
and 
\be\nn%
\sum^{s^+}_{q^+=\frac 1 2}\oplus\left|E_0,q^+\right>\oplus\sum^{s^-}_{q^-=\frac 1 2}\oplus\left|E_0,q^-\right>.%
\ee%
For simplicity we consider the case of odd $d$.
The case of even $d$ can be obtained analogously by discarding $\pm$
labels.

Using the index decomposition $n = 1,\ldots,d-1\:,\; \bullet = d$
and denoting $t^+\equiv T^{+\bullet}$, the module
$\Verma{E_0}{s}{d}$ is spanned by the vectors
\be%
T^{+n_1}\ldots T^{+n_N}(t^+)^k\left|E_0,q^\pm\right>\,,%
\ee%
where $N,k=0,1,\ldots \infty\,, \quad  q^\pm={\frac 1
2}^\pm,\ldots,s^\pm$.

Let $W_{kq}^\pm$ be a vector space spanned by%
\be\nn%
T^{+n_1}\ldots T^{+n_N}(t^+)^k\left|E_0,q^\pm\right>\:,\quad N=0,1,\ldots%
\ee%
From the definition of $W_{kq}^\pm$ it follows that
\be\label{UnitWMinus} %
T^{-n}W_{kq}^\pm\subset W_{kq}^\pm\oplus W_{k-1\:q-1}^\pm \oplus
W_{k-1\:q+1}^\pm\oplus W_{k-2\:q}^\pm\oplus
W_{k-1\:q}^\mp\:,%
\ee%
\be\label{UnitWPlus} %
T^{+n}W_{kq}^\pm\subset W_{kq}^\pm\:,\quad T^{nm}W_{kq}^\pm\subset
W_{kq}^\pm \:,\quad EW_{kq}^\pm\subset W_{kq}^\pm\,. %
\ee%

Let $V_{ij}$ be vector spaces defined recurrently as
\be\nn%
V_{ij}=\sum^j_{q=-s}\oplus U_{iq}\oplus V_{i-1\:s} \:, %
\ee%
where $j=-s,\ldots, s$ and
\[%
U_{ij}=\left\{
    \begin{array}{ll}
        W^+_{ij} &\text{if}\ j=\frac 1 2,\ldots,s\\
        W^-_{i\ -j} &\text{if}\ j=-s,\ldots,-\frac 1 2\\
    \end{array}
\right.\]%
According to this definition,
\be%
V_{ij}\subset V_{ij+1}\,,\qquad V_{ij}\subset V_{i+1\:j}\,.%
\ee%

From (\ref{UnitWMinus}) and (\ref{UnitWPlus}) it follows that
$V_{ij}$ are $o(d-1,2)$--modules. As a result we obtain the
filtration of the $o(d-1,2)$--module $\Verma{E_0}{s}{d}$ by the
$o(d-1,2)$--modules $V_{ij}$
\be\nn%
\{0\}\subset V_{0\: -s}\subset\ldots\subset V_{0\: -\frac 1
2}\subset V_{0\: \frac 1 2}\subset\ldots\subset
V_{0\: s}\subset V_{1\:-s}\subset\ldots = \Verma{E_0}{s}{d} \,.%
\ee%

Obviously, the composition factors form the following
$o(d-1,2)$--modules
\be
\begin{array}{rll}%
V_{ij}/V_{ij-1}&\simeq \Verma{E_0+i}{j^+}{d-1}\:,\ &j=\frac 1 2,\ldots,s\:,\\ %
V_{ij}/V_{ij-1}&\simeq \Verma{E_0+i}{-j^-}{d-1}\:,\ &j=-s+1,\ldots,-\frac 1 2\:,\\ %
V_{i\:-s}/V_{i-1\:s}&\simeq \Verma{E_0+i}{s^-}{d-1}\,, %
\end{array} %
\ee%
which are irreducible because the inequalities
(\ref{UnitCondLowest}), (\ref{UnitCondS}) are satisfied in $d$
dimensions as a consequence of those in $d+1$ dimensions.

$\Verma{E_0}{s}{d}$ with $E_0\geq E_0({\bf s})$ is a unitary
$o(d-1,2)$--module by the same reason as in the bosonic case. Taking
into account that a reducible unitary module is fully reducible
({\it i.e.} it decomposes into direct sum of irreducible
submodules), we obtain (\ref{dec_even_app}) and (\ref{dec_odd_app}).
\end{prf}


\end{document}